\documentclass[aos]{imsart}
\usepackage{amsmath}
\usepackage{graphicx,psfrag,epsf}
\usepackage{enumerate}
\usepackage{multirow}
\usepackage{url} 
\usepackage{xcolor}
\usepackage{amsfonts}
\usepackage{algorithm}
\usepackage{algorithmic}
\usepackage{float}
\usepackage{amsmath}
\usepackage{amssymb}
 \usepackage{multirow}

\RequirePackage{amsthm,amsmath,amsfonts,amssymb}
\RequirePackage[authoryear]{natbib}
\RequirePackage[colorlinks,citecolor=blue,urlcolor=blue]{hyperref}
\RequirePackage{graphicx}

\startlocaldefs
\theoremstyle{plain}

\newtheorem{theorem}{Theorem}[section]

\newtheorem{proposition}{Proposition}[section]

\newtheorem{corollary}{Corollary}[section]

\theoremstyle{remark}

\endlocaldefs

\begin{document}

\begin{frontmatter}
\title{Quantifying Uncertainty in Classification Performance: ROC Confidence Bands Using Conformal Prediction
}

\begin{aug}
\author[A]{\fnms{Zheshi}~\snm{Zheng}\ead[label=e1]{zszheng@umich.edu}\orcid{0000-0002-6383-434X}},
\author[A]{\fnms{Bo}~\snm{Yang}\ead[label=e2]{ybb@umich.edu}}
\and
\author[A]{\fnms{Peter}~\snm{Song}\ead[label=e3]{pxsong@umich.edu}}
\address[A]{Department of Biostatistics,
University of Michigan\printead[presep={,\ }]{e1,e2,e3}}

\end{aug}

\begin{abstract}
To evaluate a classification algorithm, it is common practice to plot the ROC curve using test data. However, the inherent randomness in the test data can undermine our confidence in the conclusions drawn from the ROC curve, necessitating uncertainty quantification. In this article, we propose an algorithm to construct confidence bands for the ROC curve, quantifying the uncertainty of classification on the test data in terms of sensitivity and specificity. The algorithm is based on a procedure called conformal prediction, which constructs individualized confidence intervals for the test set and the confidence bands for the ROC curve can be obtained by combining the individualized intervals together. Furthermore, we address both scenarios where the test data are either iid or non-iid relative to the observed data set and propose distinct algorithms for each case with valid coverage probability. The proposed method is validated through both theoretical results and numerical experiments.
\end{abstract}


\end{frontmatter}

\section{Introduction}
First developed from signal detection theory \citep{fukunaga2013introduction}, the Receiver Operating Characteristic (ROC) curve visualizes a classifier's trade-off between hit rates and false alarm rates at different cut-off points (thresholds). Recent statistical research has shown that summary statistics of the ROC curve, such as the area under the curve (AUC), are effective evaluation metrics and are widely used in fields like biomedicine \citep{kumar2011receiver} and computer science \citep{bradley1997use}.
However, another crucial component to consider when evaluating an algorithm is the confidence (or uncertainty) level regarding the performance stability of the algorithm: if a classification algorithm's ROC curves vary significantly with different test data, our confidence in the algorithm's reliability diminishes. Current research on ROC uncertainty quantification primarily focuses on the ROC curve summary statistics like AUC, using techniques such as (a) making distributional assumptions about negative and positive groups \citep{reiser1997confidence}, (b) applying asymptotic theories for empirical AUC \citep{cortes2004confidence}, or (c) using the Bootstrap method \citep{noma2021confidence}. However, AUC can lose significant information when summarizing the ROC curve, leading to criticism such as \cite{lobo2008auc}, thus inadequately addressing the issue. Additionally, the commonly used Bootstrap method, although can be applied on the ROC curve directly, often lacks theoretical guarantees and tends to underestimate the uncertainty of an algorithm, resulting in confidence bands that are too narrow for poor algorithms.
In this article, we develop algorithms to construct confidence bands for ROC curves, accurately quantifying the confidence level (uncertainty) of classification with respect to the test data.

\begin{figure}
    \centering
    \includegraphics[width = 4in]{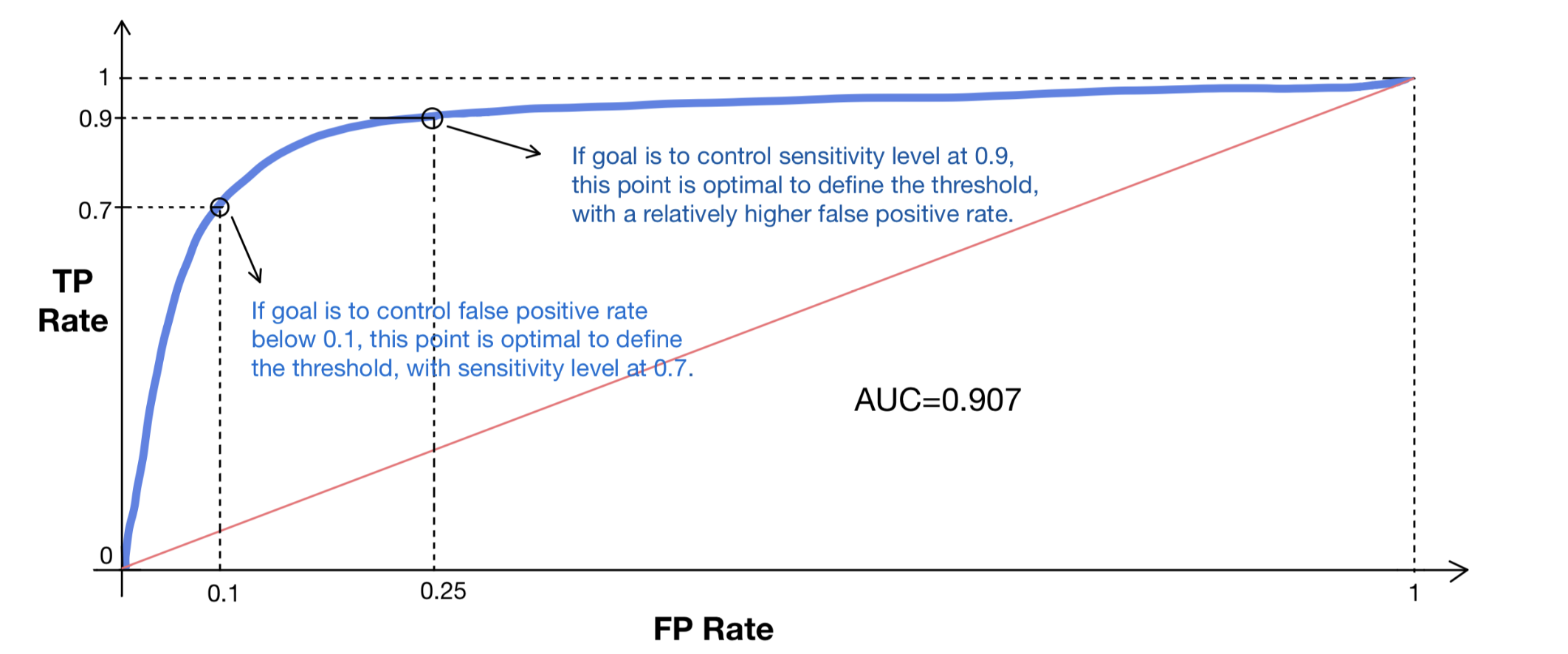}
    \caption{TP vs. FP rate at different classification thresholds.}
    \label{fig: ex of ROC}
\end{figure}

With a fixed threshold, a classification algorithm will predict each item into negative ($0$) or positive ($1$), thus we can create contingency table as shown in Table~\ref{tab: contingency}. Define sensitivity as $\text{TP Rate} = \frac{TP}{TP+FN}$ and specificity as $\text{FP Rate} = \frac{FP}{FP+TN}$, and ROC curve plots the TP Rate versus FP Rate (FPR) \citep{fawcett2006introduction}, as shown in Figure~\ref{fig: ex of ROC}. To explain why ROC curve can be used to evaluate the classification performance, we first notice that a good classification should maintains high TP Rate while control the FP Rate at a low level, thus the closer its ROC curve to the top left corner, the better classification. Thus AUC is used as a single number summary of ROC curve, which calculate the area under the ROC curve \citep{huang2005using, bradley1997use}. Regarded as the null hypothesis, if we adopt random classification that classify items into positive and negative with equal probability, the ROC curve will converge to the red line $y=x$ with AUC equals $0.5$. 

\begin{table}[htbp]
    \centering
    \caption{}
    \begin{tabular}{c|cc}
    \hline\hline
    Fixed threshold&Predicted Positive&Predict Negative\\
    \hline
    Observed Positive &True Positive (TP) & False Negative (FN)\\
    Observed Negative &False Positive (FP) & True Negative (TN)\\
    \hline\hline
    \end{tabular}
    \label{tab: contingency}
\end{table}

\begin{figure}
    \centering
    \includegraphics[width = 5in]{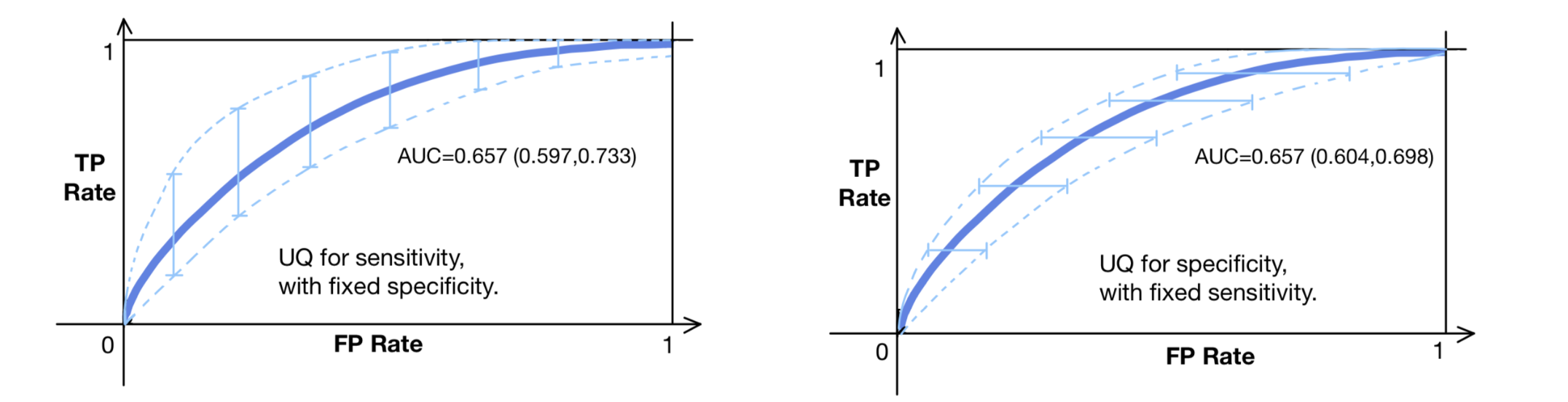}
    \caption{Examples of ROC faith bands.}
    \label{fig: ex of ROC_band}
\end{figure}

To construct confidence bands for the ROC curve, we need to construct confidence intervals for the TP Rate and FP Rate, with their uncertainties arising separately from the negative and positive groups. Therefore, we provide two types of confidence bands for the ROC curve concerning the uncertainty of sensitivity and specificity, as illustrated in Figure~\ref{fig: ex of ROC_band}. To quantify these two types of uncertainty, our approach is to construct confidence intervals for the classification of each individual in the test data with respect to sensitivity or specificity, and then aggregate these individual confidence intervals to form population-level confidence intervals for the TP Rate and FP Rate.

To construct individualized confidence intervals, we will adopt the conformal prediction approach. Introduced by Vovk in his book \cite{Vovk2005}, conformal prediction is a robust prediction inference framework. \cite{lei2018distribution} introduced split conformal and Jackknife-plus conformal procedures in a regression setting, which are widely used in statistics and machine learning literature. As discussed in \cite{xie2022homeostasis}, conformal prediction is distribution-free in the sense that even if an incorrect algorithm or model is used, the prediction interval still guarantees valid coverage, though the interval length may increase. Therefore, in our problem, if a suboptimal algorithm is chosen for classification, we can expect the confidence interval to accurately quantify uncertainty by covering the truth with high confidence, while its large length indicates large uncertainty.
Additionally, we adjust the conformal prediction algorithm to handle cases where the test data are not identically and independently distributed (iid) with the observed data set. Standard conformal prediction is known to fail when the test data are non-iid with the observed data \citep{xie2022homeostasis}. To address this issue, we adopt the iFusion idea from \cite{shen2020fusion}, using only "similar" individuals that are assumed to be iid with the test data for calibration in the conformal procedure.

In summary, we develop confidence intervals for the ROC curve to quantify the uncertainty of the classification algorithm with respect to sensitivity and specificity. Our main contributions are three-fold. First, we propose a confidence inference framework for the ROC curve that measures the uncertainty level of a classification algorithm concerning the sensitivity and specificity of the test data. To the best of our knowledge, this article is the first to develop distribution-free ROC uncertainty quantification that assesses the performance of classification on the chosen test data. This method can be particularly useful for evaluating machine learning classification algorithms without theoretical consistency. Second, we construct confidence intervals for the TP Rate and FP Rate for both iid and non-iid test data using an adjusted conformal prediction procedure. Specifically, for the non-iid setting, achieving valid coverage with conformal prediction is challenging. We address this issue by proposing a novel localization approach inspired by "iFusion learning." Third, we provide theoretical coverage guarantees for the proposed confidence intervals, supported by numerical experiments.

The rest of the paper is organized as follows. In Section 2, we define the notations for our problem and introduce the standard conformal prediction procedure. We also present the preliminary modifications to the standard procedure to ensure the algorithm fits our problem setup. In Section 3, we introduce our proposed methodology to construct confidence bands for the ROC curve. We begin with the case where the test data are iid with the observed data and then extend the methodology to the non-iid case. We also briefly discuss how to extend the method to multi-label classification problems and survival analysis. Section 4 contains all the theoretical proofs, and we include our simulation studies in Section 5. We conclude the paper with a discussion in Section 6.

\section{Preliminaries}
\subsection{Problem setup and notations}
Assume an iid data set \(\mathcal{D}_{obs} = \{(y_1,x_1),\cdots,(y_n,x_n)\}\) is observed, where \(x_i \in \mathbb{R}^d\) are numeric vectors of features and \(y_i \in \{0,1\}\) with \(y_i \sim \text{Ber}(\pi(x_i))\) and the function \(\pi(\cdot): \mathbb{R}^d \mapsto [0,1]\). We randomly split the observed data set into \(\mathcal{D}_{obs} = \mathcal{D}_{tr} \cup \mathcal{D}_{ca}\), and following standard practice, the sizes of the training and calibration sets will be proportional to each other (i.e., if \(|\mathcal{D}_{obs}| \to \infty\), then both \(|\mathcal{D}_{tr}|\) and \(|\mathcal{D}_{ca}|\) also \(\to \infty\) at the same rate of \(n\)). 
\(\mathcal{D}_{tr}\) is used to train an algorithm to obtain \(p_n(\cdot) = p_n(\cdot|\mathcal{D}_{tr})\) to estimate \(\pi(\cdot)\), and \(\mathcal{D}_{ca}\) is used to calibrate \(p_n(\cdot)\) and quantify its uncertainty. An iid data set \(\mathcal{D}_{tst} = \{(y_j,x_j): j \in \mathcal{I}_{tst}\}\) is used to test the performance of \(p_n(\cdot)\) and plot its ROC curve. We assume \(\mathcal{D}_{tst}\) and \(\mathcal{D}_{obs}\) are independent, with \(y_{n+j} \sim \text{Ber}(\pi(x_{n+j}))\). We will consider both cases where \(\mathcal{D}_{tst}\) is from the same or a different population as \(\mathcal{D}_{obs}\). For \(\mathcal{D}_{tst}\) from the same population as \(\mathcal{D}_{obs}\), we assume:

\begin{itemize}
    \item[(A1)] For each \(j=1,2,\cdots,m\), \(x_{n+j}\) is iid sampled from the same population as \(\{x_i\}_{i=1}^n\).
\end{itemize}

Denote the index set of \(\mathcal{D}_{obs}, \mathcal{D}_{tr}, \mathcal{D}_{ca}\) as \(\mathcal{I}_{obs}, \mathcal{I}_{tr}, \mathcal{I}_{ca}\). Denote \(\mathcal{D}_{tst}^k = \{x_i : i \in \mathcal{I}_{tst}, y_i = k\}\) and their corresponding index set as \(\mathcal{I}_{tst}^k\), for \(k=0,1\). Given any threshold \(\lambda \in [0,1]\), false positive rate (FPR) and true positive rate (TPR) are defined as:
\begin{align*}
    &TPR(\lambda) = \frac{\text{\#detected positives among observed positives}}{\text{\#observed positives}} = \frac{\sum_{j \in \mathcal{I}_{tst}^1} \mathbf{1}(p(x_j) \ge \lambda)}{|\mathcal{D}_{tst}^1|}\\
    &FPR(\lambda) = \frac{\text{\#detected positives among observed negatives}}{\text{\#observed negatives}} = \frac{\sum_{j \in \mathcal{I}_{tst}^0} \mathbf{1}(p(x_j) \ge \lambda)}{|\mathcal{D}_{tst}^0|}
\end{align*}
where \(\mathbf{1}(A)\) is the indicator function that takes the value 1 if \(A\) is true, and 0 otherwise. We can further define the true positive fraction (TPF) \(Se_p(\lambda) = \Pr(p_n(x_j) > \lambda \mid j \in \mathcal{I}_{tst}^1)\) and the false positive fraction (FPF) \(Sp_p(\lambda) = \Pr(p_n(x_j) \le \lambda \mid j \in \mathcal{I}_{tst}^0)\), and we know that \(TPR(\lambda) \to Se_p(\lambda)\) a.s. as \(|\mathcal{D}_{tst}^1| \to \infty\) and \(FPR(\lambda) \to Sp_p(\lambda)\) a.s. as \(|\mathcal{D}_{tst}^0| \to \infty\).

To quantify the uncertainty of \(p_n(\cdot)\), our idea is to obtain individualized conformal prediction for each item in \(\mathcal{D}_{tst}\), and then transform it into a confidence band for the ROC curve. Thus, we will regard each \(x_{n+j}\) in the testing set as a new object, and index it by \((x_{new}, y_{new})\).

Our individualized confidence interval should have the coverage probability conditioned on the positive label (for inference of \(TPR(\lambda)\)) or negative label (for inference of \(FPR(\lambda)\)). Thus, our prediction will target covering the following random variables:
\begin{align*}
    &p_{new}^{(k)} = p_n(x_{new} \mid y_{new} = k) \text{ and } \pi_{new}^{(k)} = \pi(x_{new} \mid y_{new} = k), k = 0,1.
\end{align*}
Here \(p_{new}^{(k)}\) and \(\pi_{new}^{(k)}\) are random variables with randomness from \(x_{new}\), and in addition, \(p_{new}^{(k)}\) has extra randomness from \(\mathcal{D}_{tr}\) as the function \(p_n(\cdot)\) is based on the training data set. \(1 - Se_p(\cdot)\) is the distribution function for \(p_{new}^{(1)}\) and \(Sp_p(\cdot)\) is the distribution function for \(p_{new}^{(0)}\). If we further define the oracle TPF and FPF for \(\pi(\cdot)\) (not observed in practice) to be 
\begin{align*}
    &TPR^{(o)}(\lambda) = \frac{\sum_{j \in \mathcal{I}_{tst}^1} \mathbf{1}(\pi(x_j) \ge \lambda)}{|\mathcal{D}_{tst}^1|};\quad FPR^{(o)}(\lambda) = \frac{\sum_{j \in \mathcal{I}_{tst}^0} \mathbf{1}(\pi(x_j) \ge \lambda)}{|\mathcal{D}_{tst}^0|}\\
    &Se_{\pi}(\lambda) = \Pr(\pi(x_j) > \lambda \mid j \in \mathcal{I}_{tst}^1);\quad Sp_{\pi}(\lambda) = \Pr(\pi(x_j) \le \lambda \mid j \in \mathcal{I}_{tst}^0)
\end{align*}
then the distribution function for \(\pi_{new}^{(1)}\) is \(1 - Se_{\pi}(\cdot)\) and the distribution function for \(\pi_{new}^{(0)}\) is \(Sp_{\pi}(\cdot)\). With the defined target, we will develop confidence intervals that guarantee the following coverage probabilities:
{\scriptsize
\begin{align*}
    &\Pr(\pi_n(x_{new}) \in CI^{(k)}(x_{new}, \alpha) \mid y_{new} = k) = \Pr(\pi_{new}^{(k)} \in CI^{(k)}(x_{new}, \alpha) \mid y_{new} = k) \ge 1 - \alpha - o(1)\\
    &\Pr(p_n(x_{new}) \in CI^{(k)}(x_{new}, \alpha) \mid y_{new} = k) = \Pr(p_{new}^{(k)} \in CI^{(k)}(x_{new}, \alpha) \mid y_{new} = k) \ge 1 - o(1)
\end{align*}}

\subsection{Prediction interval for $\pi(x_{new})$ with conformal prediction}
Before constructing confidence intervals for \( p_{new}^{(k)} \) and \( \pi_{new}^{(k)} \), we first construct the projection interval for \(\pi(x_{new})\). This differs from standard conformal prediction, which directly targets predicting \( y_{new} \). However, we can start with the case where we observe \(\pi(x_i)\) for \( i \in \mathcal{I}_{obs} \). Using the standard conformal prediction procedure, define conformity scores \( R_i = L_{\pi}(x_i) - L_p(x_i) \) for \( i \in \mathcal{I}_{ca} \), where \( L_{\pi}(x_i) = \text{logit}(\pi(x_i)) = \ln \frac{\pi(x_i)}{1-\pi(x_i)} \) and \( L_p(x_i) = \text{logit}(p_n(x_i)) \). Similarly, for any potential value \(\pi\) of \(\pi(x_{new})\), we define \( R(\pi) = \text{logit}(\pi) - L_p(x_{new}) \). We use the logit function here to ensure that the probability bound will always lie between \([0,1]\). Notice that with Assumption (A1), \( R_i \) and \( R(\pi(x_{new})) \) are exchangeable. Thus, we can define the p-value of the potential value \(\pi\) as \( p(\pi) = \frac{1 + \sum_{i \in \mathcal{I}_{ca}} \mathbf{1}(R(\pi) \ge R_i)}{|\mathcal{D}_{ca}| + 1} \). The prediction interval can then be obtained from
\begin{align*}
    &PI^{std}(x_{new}, \alpha) = \{\pi \in [0,1] : 2 \min(p(\pi), 1 - p(\pi)) \ge \alpha\} \\
    &\quad\quad= \left[ \text{expit}\left\{L_p(x_{new}) + q_{\alpha/2}(\{R_i\}_{i \in \mathcal{I}_{ca}})\right\}, \text{expit}\left\{L_p(x_{new}) + q_{1-\alpha/2}(\{R_i\}_{i \in \mathcal{I}_{ca}})\right\} \right]
\end{align*}
where \(\text{expit}(x) = \text{logit}^{-1}(x) = \frac{1}{1 + \exp(-x)}\), and \( q_{\alpha/2}(\mathcal{A}) \) is the \(\left([\alpha(|\mathcal{A}| + 1)/2] - 1\right)\)-th order statistic for the set \(\mathcal{A}\) and \( q_{1-\alpha/2}(\mathcal{A}) \) is the \([(1 - \alpha/2)(|\mathcal{A}| + 1)]\)-th order statistic for the set \(\mathcal{A}\), with \([a]\) being the largest integer that does not exceed \(a\). Without loss of generality, we will assume \( R_i \) are continuous with the following assumption:
\begin{itemize}
    \item[(A2)] Let \( F_{R}(\cdot) \) denote the CDF of \( R_i \). For any \( r, r_0 \) in the support of function \( F_R \), there exists a constant \( C > 0 \) such that \( |F_R(r) - F_R(r_0)| \le C|r - r_0| \).
\end{itemize}

\begin{proposition}
Assume (A1)-(A2) hold, and define the probability space to be \(\big(\sigma(\mathcal{D}_{obs} \cup (x_{new}, y_{new})), P\big)\), where \(\sigma(\mathcal{A})\) is the \(\sigma\)-algebra for the random variable set \(\mathcal{A}\). Then we have
\begin{equation*}
    P(\pi_{new} \in PI^{std}(x_{new}, \alpha)) \ge 1 - \alpha.
\end{equation*}
\end{proposition}
The proof of this proposition is standard and based on the order statistics of an exchangeable set.

However, in practice, we do not observe \(\pi(x_i)\), and we need to use their approximations \(\tilde \pi(x_i)\) instead. A commonly used non-parametric approximation can be defined as the kernel estimator 
\begin{equation}
    \tilde \pi(x_i) = \frac{\sum_{j \in \mathcal{I}_{tr}} \mathbf{1}(y_j = 1) K_h(x_j, x_i)}{\sum_{j \in \mathcal{I}_{tr}} K_h(x_j, x_i)}.\label{eq:tilde_pi}
\end{equation}

where \( K_h(\cdot, \cdot) \) is a well-defined kernel function that acts as a similarity function. Throughout the paper, we will assume that \(\tilde \pi(x_i)\) is a good estimator for \(\pi(x_i)\) such that they satisfy the following assumption:
\begin{itemize}
    \item[(A3)] There exists \(\delta_{n} \to 0\) as \(|\mathcal{D}_{tr}| \to \infty\), such that \\
    \( \lim_{|\mathcal{D}_{ca}|\to\infty} \mathbb{E}_{x_i \sim \mathcal{D}_{ca}} \mathbf{1}(|L_{\tilde \pi}(x_i) - L_{\pi}(x_i)| > \delta_n) = 0\), where \( L_{\tilde \pi}(x_i) = \text{logit}(\tilde \pi(x_i)) \).
\end{itemize}

Define the conformity score \(\tilde{R}_i = L_{\tilde{\pi}}(x_i) - L_p(x_i)\). We obtain the prediction interval $PI(x_{new}, \alpha)$ as
\begin{align*}
\Big[ &\text{expit}\Big\{L_p(x_{new}) + q_{\alpha/2}(\{\tilde{R}_i\}_{i \in \mathcal{I}_{ca}})\Big\},\text{expit}\Big\{L_p(x_{new}) + q_{1-\alpha/2}(\{\tilde{R}_i\}_{i \in \mathcal{I}_{ca}})\Big\} \Big]
\end{align*}
The theoretical coverage guarantee for this prediction interval is given in the following theorem.
\begin{theorem}
Assume (A1) and (A2) hold, and define the same probability space as in Proposition 2.1. Then we have
\begin{equation*}
    \lim_{|\mathcal{D}_{ca}|\to\infty} P\left(\pi_{new} \in PI(x_{new}, \alpha)\right) \ge 1 - \alpha - 2C\delta_n
\end{equation*}
\end{theorem}

From the above theorem, we know that the coverage rate relates to the convergence rate of the estimator \(\tilde{\pi}(\cdot)\).

\section{Methodology}\label{sec: methodology}In this section, we will construct confidence intervals for \( p_{new}^{(k)} \) and \( \pi_{new}^{(k)} \) for \( k=0,1 \), and then transform them into confidence bands for the ROC curve. We begin with the case where \( x_{new} \) is an iid sample from the observed population and then extend our approach to the non-iid case. Furthermore, we extend our method to address multi-label classification problems.

\subsection{IID test data set}
To construct the confidence intervals, we define the conformity scores in the same way as before: \( \tilde{R}_i = L_{\tilde{\pi}}(x_i) - L_p(x_i) \). We split \(\mathcal{D}_{ca} = \mathcal{D}_{ca}^1 \cup \mathcal{D}_{ca}^0\) by \( y_i = 1 \) or \( y_i = 0 \) and denote their index sets as \(\mathcal{I}_{ca}^1\) and \(\mathcal{I}_{ca}^0\).

We will use modified versions of assumptions (A2) and (A3) as follows:
\begin{itemize}
    \item[(A2')] Let \( F_{R,k}(\cdot) \) denote the CDF of \(\{R_i\}_{i \in \mathcal{I}_{ca}^k}\). For any \( x, x_0 \) in the support of \( F_{R,k} \), there exists a constant \( C_k > 0 \) such that \( |F_{R,k}(x) - F_{R,k}(x_0)| \le C_k |x - x_0| \), for \( k=1,0 \).
\end{itemize}

\begin{itemize}
    \item[(A3')] There exist \( \delta_{n}^{(k)} \to 0 \) as \( |\mathcal{D}_{tr}| \to \infty \), such that \( \lim_{\substack{|\mathcal{D}_{ca}| \to \infty}} \mathbb{E}_{x_i \sim \mathcal{D}_{ca}^k} \mathbf{1}\big(|L_{\tilde{\pi}}(x_i) - L_{\pi}(x_i)| > \delta_{n}^{(k)}\big) = 0 \), for \( k=1,0 \).
\end{itemize}

Given \( y_{new} = k \), \(\mathcal{D}_{ca}^k\) and \((x_{new}, y_{new})\) are iid, then the confidence interval \( CI(x_{new}, \alpha;\)
\(y_{new}\) \( = k) \) for \(\pi(x_{new})\) given \( y_{new} = k \) can be obtained by
\begin{align*}
\left[\text{expit}\left\{L_p(x_{new}) + q_{\alpha/2}(\{\tilde{R}_i\}_{i \in \mathcal{I}_{ca}^k})\right\}, \text{expit}\left\{L_p(x_{new}) + q_{1-\alpha/2}(\{\tilde{R}_i\}_{i \in \mathcal{I}_{ca}^k})\right\}\right]
\end{align*}
The coverage of \( CI(x_{new}, \alpha; y_{new} = k) \) is guaranteed by the following Theorem~\ref{thm: individual credible interval}.
\begin{theorem}\label{thm: individual credible interval}
Assume (A1), (A2'), and (A3') hold and define the same probability space as in Proposition 2.1. Then, we have that for \( k=0,1 \),
\begin{equation}\label{eq: individual random uncertainty}
    \lim_{|\mathcal{D}_{ca}^k| \to \infty} P\left(\pi(x_{new}) \in CI(x_{new}, \alpha; y_{new} = k) \mid y_{new} = k\right) \ge 1 - \alpha - C_k \delta_{n}^{(k)}
\end{equation}
In addition, if we further assume that \( F_{R,k}^{-1}(\alpha/2) < 0 \) and \( F_{R,k}^{-1}(1 - \alpha/2) > 0 \), then
\begin{equation}\label{eq: individual coverage for trained pi}
    \lim_{|\mathcal{D}_{tr}|, |\mathcal{D}_{ca}^k| \to \infty} P\left(p_n(x_{new}) \in CI(x_{new}, \alpha; y_{new} = k) \mid y_{new} = k\right) = 1.
\end{equation}
\end{theorem}

In the above theorem, (\ref{eq: individual random uncertainty}) shows that the confidence interval correctly quantifies the uncertainty of the algorithm for each individual in the testing set, and (\ref{eq: individual coverage for trained pi}) shows the monotonicity between the prediction given by the algorithm and the bounds of the credible intervals, thus guaranteeing that the ROC curve band will cover the ROC plotted by \( p_n(\cdot) \).

Having developed the confidence interval for each subject in the testing set, we then transform it into the confidence band for the ROC curves. The detailed algorithm is described below:

\begin{algorithm}[H]\label{alg:iid FP_TP}
\caption{Confidence interval for $FP(\lambda)$ and $TP(\lambda)$ under iid setting.}
\begin{algorithmic}[1]
\REQUIRE Dataset $\mathcal{D}_{obs}$, $\mathcal{D}_{tst}$, significance level $\alpha$, training algorithm $p_n(\cdot)$.
\STATE Split $\mathcal{D}_{obs} = \mathcal{D}_{tr} \cup \mathcal{D}_{ca}$, train the model based on $\mathcal{D}_{tr}$, and obtain $p_n(\cdot) = p_n(\cdot \mid \mathcal{D}_{tr})$.
\FOR{each instance $(x_{j}, y_{j})$ in $\mathcal{D}_{tst}$}
\STATE Calculate \(\tilde{R}_i = L_{\tilde{\pi}}(x_i) - L_p(x_i)\), for \(i \in \mathcal{I}_{ca}^k\) and \(k = y_{new}\). \(\tilde{\pi}(x_i)\) is defined in (\ref{eq:tilde_pi}).
\STATE Calculate \(CI(x_j, \alpha; y_j = k) = [b_{lo}(x_j, \alpha; k), b_{up}(x_j, \alpha; k)]\), with
\vspace{-2mm}
\begin{align*}
    &b_{lo}(x_j, \alpha; k) = \text{expit}\big\{L_p(x_j) + q_{\alpha/2}(\{\tilde{R}_i\}_{i \in \mathcal{I}_{ca}^k})\big\},\\
    &b_{up}(x_j, \alpha; k) = \text{expit}\big\{L_p(x_j) + q_{1-\alpha/2}(\{\tilde{R}_i\}_{i \in \mathcal{I}_{ca}^k})\big\}.
\end{align*}
\vspace{-3mm}
\ENDFOR
\STATE Obtain the confidence interval for $TPR(\lambda)$ and $FPR(\lambda)$ from
\vspace{-2mm}
{\scriptsize
\begin{align*}
    &CI_{\lambda}^{sen}(\mathcal{D}_{tst}^1, \alpha) = \bigg[\frac{1}{|\mathcal{I}_{tst}^1|}\sum_{j \in \mathcal{I}_{tst}^1} \mathbf{1}\big(b_{lo}(x_j, \alpha; 1) > \lambda\big),\quad \frac{1}{|\mathcal{I}_{tst}^1|}\sum_{j \in \mathcal{I}_{tst}^1} \mathbf{1}\big(b_{up}(x_j, \alpha; 1) > \lambda\big)\bigg]\\
    &CI_{\lambda}^{spe}(\mathcal{D}_{tst}^0, \alpha) = \bigg[\frac{1}{|\mathcal{I}_{tst}^0|}\sum_{j \in \mathcal{I}_{tst}^0} \mathbf{1}\big(b_{lo}(x_j, \alpha; 0) > \lambda\big),\quad \frac{1}{|\mathcal{I}_{tst}^0|}\sum_{j \in \mathcal{I}_{tst}^0} \mathbf{1}\big(b_{up}(x_j, \alpha; 0) > \lambda\big)\bigg]
\end{align*}
}
\vspace{-2mm}
\RETURN \(\left(CI_{\lambda}^{sen}(\mathcal{D}_{tst}^1, \alpha), CI_{\lambda}^{spe}(\mathcal{D}_{tst}^0, \alpha)\right) \).
\end{algorithmic}
\end{algorithm}

In practice, to make the splitting procedure more stable, we can move line 1 below line 2 inside the ``for" loop in Algorithm~\ref{alg:iid FP_TP}. In this case, we can have variance for the individual confidence interval \(CI(x_j, \alpha; y_j = k)\) due to the splitting randomness. The empirical performance will not change significantly.



To analyze the theoretical property of  $CI_{\lambda}^{sen}(\mathcal{D}_{tst}^1,\alpha)$ and $CI_{\lambda}^{spe}(\mathcal{D}_{tst}^0,\alpha)$, we want to show that (1) for uncertainty quantification, they will cover the oracle ROC curve (given by $\pi(x_{new})$, $x_{new}\in\mathcal{D}_{tst}$) with confidence level $1-\alpha$, and (2) for valid definition of curve band, they will almost surely cover the ROC curve (given by $p_n(x_{new})$, $x_{new}\in\mathcal{D}_{tst}$). Thus, we develop the following Theorem~\ref{thm: FP,TP}.

\begin{theorem}\label{thm: FP,TP}
Assume $\mathcal{D}_{ca}$ and $\mathcal{D}_{tr}$ are exchangeable, and the assumptions in Theorem 2.2 still holds. Define the probability space to be $\big(\sigma\big(\mathcal{D}_{obs}\cup\mathcal{D}_{tst}\big),P\big)$, we randomly choose $\lambda = \pi(x_s)$, $s \in\mathcal{I}_{tst}$, then
\begin{align*}
    &\lim_{|\mathcal{D}_{tr}|,|\mathcal{D}_{ca}^1|\to\infty}P\big(TPR^{(o)}\big(\pi(x_s)\big) \in CI_{\lambda}^{sen}(\mathcal{D}_{tst}^1,\alpha)\big)\ge 1-\alpha,\text{ for $s\in\mathcal{I}_{tst}^1$,}\\
    &\lim_{|\mathcal{D}_{tr}|,|\mathcal{D}_{ca}^0|\to\infty}P\big(FPR^{(o)}\big(\pi(x_s)\big) \in CI_{\lambda}^{spe}(\mathcal{D}_{tst}^0,\alpha)\big)\ge 1-\alpha,\text{ for $s\in\mathcal{I}_{tst}^0$,}
\end{align*}
And in addition, if we further assume that $F_{R,k}^{-1}(\alpha/2)<0$ and $F_{R,k}^{-1}(1-\alpha/2)>0$, then for any $\lambda\in(0,1)$, almost surely,
\begin{align*}
    &TPR(\lambda) \in CI_{\lambda}^{sen}(\mathcal{D}_{tst}^1,\alpha),\quad FPR(\lambda) \in CI_{\lambda}^{spe}(\mathcal{D}_{tst}^0,\alpha)
\end{align*}
\end{theorem}
Notice that by definition, the ROC curve is a step function with jumps at \(\lambda = \pi(x_s)\), \( s \in \mathcal{I}_{tst} \). Thus, the first part of the theorem above shows that if we randomly choose a jump point on the ROC, the confidence interval with fixed sensitivity (for a point with a negative outcome) or with sensitivity (for a point with a positive outcome) will cover the oracle \( FPR \) or \( TPR \) with confidence level converging to \( 1 - \alpha \).
Detailed proof of Theorem~\ref{thm: FP,TP} is given in the following section.

\subsection{Non-iid test data set}
In practice, test data can come from a different population on which we want to evaluate the algorithm's performance. For example, algorithms can be trained on a large and comprehensive but unavailable data set, and we may only test it on a small population that is inconclusive. In this case, we need to generalize the previous method to obtain the conditional confidence interval \( CI_c(x_{new}, \alpha, y_{new} = k) \) that specifically captures the uncertainty for the specific population. However, as discussed in \cite{foygel2021limits}, such conditional inference is impossible without further assumptions. Our idea is to create a subgroup of the calibration set \(\mathcal{N}_{new} = \{(x_i, y_i) \in \mathcal{D}_{ca} : \text{dist}(x_{new}, x_i) \le d\}\) consisting of individuals that are similar to our prediction target. We further assume \((x_{new}, y_{new})\) and \(\mathcal{N}_{new}\) are iid under the following assumption (A4):
\begin{itemize}
    \item[(A4)] $(x_{new},y_{new})$ and $\mathcal{N}_{new}$ are iid from the same distribution.
\end{itemize}



The detailed implementation is introduced in the following algorithm. 
\begin{algorithm}[H]
\caption{Conformal faith interval for $FP(\lambda)$ and $TP(\lambda)$ under non-iid setting.}
\begin{algorithmic}[1]
\REQUIRE Dataset $\mathcal{D}_{obs}$, $\mathcal{D}_{tst}$, significance level $\alpha$, training algorithm $p_n(\cdot)$, distance function $dist(\cdot,\cdot)$ and neighbor size $N$.
\FOR{each instance $(x_{j}, y_{j})$ in $D_{tst}$}
\STATE Obtain $\mathcal{N}_{j} = \{(x_k,y_k)\in\mathcal{D}_{ca}:  dist(x_k,x_{j})<d\}$, choose the smallest $d$ such that $|\mathcal{N}_{j}| \ge N$.
\STATE Calculate $\tilde R_i = L_{\tilde\pi}(x_i) - L_{p}(x_i)$, for $i:(x_i,y_i)\in\mathcal{N}_j$ and $k=y_{new}$. $\tilde\pi(x_i)$ defined in (\ref{eq:tilde_pi}).
\STATE Calculate $CI_c(x_j,\alpha;y_j=k) = [b_{lo}(x_j,\alpha;k),b_{up}(x_j,\alpha;k)] $, with
\vspace{-2mm}
\begin{align*}\small
    &b_{c,lo}(x_j,\alpha;k) =expit\big\{L_{\tilde\pi}(x_j) +q_{\alpha/2}(\{\tilde R_i\}_{i:(x_i,y_i)\in\mathcal{N}_{j},y_i=k})\big\}\\
    &b_{c,up}(x_j,\alpha;k)=expit\big\{L_{\tilde\pi}(x_j) +q_{1-\alpha/2}(\{\tilde R_i\}_{i:(x_i,y_i)\in\mathcal{N}_{j},y_i=k})\big\}
\end{align*}
\ENDFOR
\STATE Obtain faith interval for $TPR(\lambda)$ and $FPR(\lambda)$ from
\vspace{-2mm}
{\scriptsize
\begin{align*}
    &CI_{c,\lambda}^{sen}(\mathcal{D}_{tst}^1,\alpha) = \bigg[\frac{1}{|\mathcal{I}_{tst}^1|}\sum_{j\in\mathcal{I}_{tst}^1} {\bf 1}\big(b_{c,lo}(x_j,\alpha;1)>\lambda\big),\quad \frac{1}{|\mathcal{I}_{tst}^1|}\sum_{j\in\mathcal{I}_{tst}^1} {\bf 1}\big(b_{c,up}(x_j,\alpha;1)>\lambda\big) \bigg]\\
    &CI_{c,\lambda}^{spe}(\mathcal{D}_{tst}^0,\alpha) = \bigg[\frac{1}{|\mathcal{I}_{tst}^0|}\sum_{j\in\mathcal{I}_{tst}^0} {\bf 1}\big(b_{c,lo}(x_j,\alpha;0)>\lambda\big),\quad \frac{1}{|\mathcal{I}_{tst}^0|}\sum_{j\in\mathcal{I}_{tst}^0} {\bf 1}\big(b_{c,up}(x_j,\alpha;0)>\lambda\big) \bigg]
\end{align*}}\\
\vspace{-2mm}
\RETURN Collect $\big\{ \big(CI_{c,\lambda}^{sen}(\mathcal{D}_{tst}^1,\alpha),CI_{c,\lambda}^{spe}(\mathcal{D}_{tst}^0,\alpha)\big):\lambda\in[0,1]\big\}$ and plot the ROC curve bands. 
\end{algorithmic}
\end{algorithm}

Similar as the iid section, we define $\mathcal{N}_{new}^k = \{(x_i,y_i)\in\mathcal{N}_{new},y_i=k\}$ and denote their index sets to be $\mathcal{I}_{new}^k$, and adjust the assumptions in the iid setting in the following:
\begin{itemize}
    \item[(A2'')] Let $F_{c,k}(\cdot)$ denotes the CDF of $\{R_i\}_{i\in\mathcal{I}_{new}^k}$, assume $F_{c,k}(\cdot)$ is Lipschitz continuous with constant $C_k'$, for $k=1,0$.
\end{itemize}
\begin{itemize}
    \item[(A3'')] There exist $\delta_{c,n}^{(k)}\to 0$ as $|\mathcal{D}_{tr}|\to \infty$, such that $\lim_{|\mathcal{N}_{new}^k|\to\infty}\mathbb E_{x_i\sim\mathcal{N}_{new}^k}{\bf 1}\big(|L_{\tilde\pi}(x_i)-L_{\pi}(x_i)|>\delta_{c,n}^{(k)}\big)=0$, for $k=1,0$.
\end{itemize}

In the following Corollary~\ref{cor: cond individual}, we develop the coverage bound for individual confidence intervals $CI_c(x_{new},\alpha,y_{new}=k)$ (proof omitted).

\begin{corollary}\label{cor: cond individual}
Assume (A4), (A2'') and (A3''),
if $|\mathcal{D}_{tr}|\to\infty$,
\begin{equation}\label{eq: individual random uncertaity non-iid}
    \lim_{|\mathcal{N}_{new}^k|\to\infty}P\bigg(\pi(x_{new})\in CI_c(x_{new},\alpha;y_{new}=k)\bigg|x_{new}, y_{new}=k\bigg)\ge 1-\alpha - C_k'\delta_{c,n}^{(k)}
\end{equation}
In addition, if we further assume that $F_{c,k}^{-1}(\alpha/2)<0$ and $F_{c,k}^{-1}(1-\alpha/2)>0$, then
\begin{equation}\label{eq: individual phat cov non-iid}
    \lim_{|\mathcal{D}_{tr}|,|\mathcal{N}_{new}^k|\to\infty}P\bigg(p_n(x_{new})\in CI_c(x_{new},\alpha;y_{new}=k)\bigg|x_{new}, y_{new}=k\bigg) = 1
\end{equation}
\end{corollary}

Then we can prove validity of confidence intervals $CI_{c,\lambda}^{sen}(\mathcal{D}_{tst}^1,\alpha) $ and $CI_{c,\lambda}^{spe}(\mathcal{D}_{tst}^0,\alpha) $ for $TPR(\lambda)$ and $FPR(\lambda)$ in the following Corollary~\ref{cor:FPR_TPR_cond} (proof omitted).

\begin{corollary}\label{cor:FPR_TPR_cond}
    Assume $\mathcal{D}_{tst}$ are iid, and assumptions (A4), (A2'') and (A3'') holds. Define the probability space to be $\big(\sigma\big(\mathcal{D}_{obs}\cup\mathcal{D}_{tst}\big),P\big)$, we randomly choose $\lambda = \pi(x_s)$, $s \in\mathcal{I}_{tst}$, then
\begin{align*}
    &\lim_{|\mathcal{D}_{tr}|\to\infty,N_{tst}^1\to\infty}P\big(TPR^{(o)}\big(\pi(x_s)\big) \in CI_{c,\lambda}^{sen}(\mathcal{D}_{tst}^1,\alpha)\big)\ge 1-\alpha,\text{ for $s\in\mathcal{I}_{tst}^1$,}\\
    &\lim_{|\mathcal{D}_{tr}|\to\infty,N_{tst}^0\to\infty}P\big(FPR^{(o)}\big(\pi(x_s)\big) \in CI_{c,\lambda}^{spe}(\mathcal{D}_{tst}^0,\alpha)\big)\ge 1-\alpha,\text{ for $s\in\mathcal{I}_{tst}^0$,}
\end{align*}
where $N_{tst}^k = \inf_{j\in\mathcal{I}_{tst}^k}\big\{|\mathcal{N}_j|\big\}$. And in addition, if the conditions in the second part of Theorem 3.1 holds, we have that for any $\lambda\in(0,1)$, almost surely,
\begin{align*}
    &TPR(\lambda) \in CI_{c,\lambda}^{sen}(\mathcal{D}_{tst}^1,\alpha),\quad FPR(\lambda) \in CI_{c,\lambda}^{spe}(\mathcal{D}_{tst}^0,\alpha)
\end{align*}
\end{corollary}

\subsection{Extension to the multi-label classification uncertainty quantification}
In practice, the subjects can be divided into multiple groups, and we may have multi-label outcomes rather than binary outcomes. In this case, the observed and test data are denoted as \(\mathcal{D}_{obs} = \{(x_i, y_i) : i \in \mathcal{I}_{obs}\}\), \( y_i \in \{1,2,\cdots,L\} \) and \(\mathcal{D}_{tst} = \{(x_{n+j}, y_{n+j}) : j \in \mathcal{I}_{tst}\} \). Here, \( x_i \) are the covariates, and the PMF of \( y_i \) is given by \(\pi(x) = \prod_{l=1}^L \pi_l(x)^{{\bf 1}(y_i = l)}\) with \(\sum_{l=1}^L \pi_l(x) = 1\). A classification algorithm will typically output some \( p_{n,l}(x_i) \) for \( l = 1, \cdots, L \), and an estimator of \( y_i \) will be \(\hat{y}_i = \arg\max_l \{p_{n,l}(x_i)\}\).

With a multi-label outcome, a common practice is to plot the ROC curve for each potential label separately. For one label \( l \in \{1,2,\cdots,L\} \), we consider the binary outcome \( y_{l,j} = {\bf 1}(y_j = l) \), and plot the ROC for \( p_{n,l}(\cdot) \). Thus, the \( TPR_l(\lambda) \) and \( FPR_l(\lambda) \) can be defined as
\begin{equation}
    TPR_l(\lambda) = \frac{1}{|\mathcal{D}_{tst}^l|}\sum_{j \in \mathcal{I}_{tst}^l} {\bf 1}\big(p_{n,l}(x_j) > \lambda\big);\quad FPR_l(\lambda) = \frac{1}{|\overline{\mathcal{D}_{tst}^l}|}\sum_{j \notin \mathcal{I}_{tst}^l} {\bf 1}\big(p_{n,l}(x_j) > \lambda\big);
\end{equation}
where \(\mathcal{D}_{tst}^l = \{(x_i, y_i) \in \mathcal{D}_{tst} : y_i = l\}\) and \(\overline{\mathcal{D}_{tst}^l} = \{(x_i, y_i) \in \mathcal{D}_{tst} : y_i \neq l\}\). Similarly, the oracle \( TPR_l^{(o)}(\lambda) \) and \( FPR_l^{(o)}(\lambda) \) can be defined accordingly for \(\pi_l(x_j)\), \( j \in \mathcal{I}_{tst} \).

To apply the previous algorithm, we first split the data \(\mathcal{D}_{obs} = \mathcal{D}_{tr} \cup \mathcal{D}_{ca}\) and define the conformity score \(\tilde{R}_i = L_{\tilde{\pi}_l}(x_i) - L_{p_l}(x_i)\), for \( i \in \mathcal{I}_{ca}^l \) if \( y_i = l \) and \( i \notin \mathcal{I}_{ca}^l \) if \( y_i \neq l \). Here, \( L_{\tilde{\pi}_l}(x_i) = \text{logit}(\tilde{\pi}_l(x_i)) \) and \( L_{p_l}(x_i) = \text{logit}(p_{n,l}(x_i)) \), with \(\tilde{\pi}_l\) defined similarly as in Equation~(\ref{eq:tilde_pi}).

\begin{algorithm}[H]
\caption{Faith interval for $FPR_l(\lambda)$ and $TPR_l(\lambda)$ for multi-label classification.}
\begin{algorithmic}[1]
\REQUIRE Dataset $\mathcal{D}_{obs}$, $\mathcal{D}_{tst}$, significance level $\alpha$, training algorithm $p_{n,l}(\cdot)$.
\FOR{each instance $(x_{j}, y_{j})$ in $D_{tst}$}
\STATE Split $\mathcal{D}_{obs} = \mathcal{D}_{tr}\cup \mathcal{D}_{ca}$, and train model based on $\mathcal{D}_{tr}$ and obtain $p_n(\cdot) = p_{n,l}(\cdot\mid \mathcal{D}_{tr})$.
\STATE Calculate $\tilde R_i = L_{\tilde\pi_l}(x_i) - L_{p_l}(x_i)$, for $i\in\mathcal{I}_{ca}^l$ if $y_i=l$, otherwise for $i\notin\mathcal{I}_{ca}^l$.
\STATE Calculate $CI(x_j,\alpha;y_j=k) = [b_{lo}(x_j,\alpha;k),b_{up}(x_j,\alpha;k)] $, with
\vspace{-2mm}
\begin{align*}
    &b_{lo}(x_j,\alpha;y_j=l) =expit\big\{L_p(x_j) +q_{\frac{\alpha}{2}}(\{\tilde R_i\}_{i\in\mathcal{I}_{ca}^l})\big\},\\
    &b_{lo}(x_j,\alpha;y_j\neq l) =expit\big\{L_p(x_j) +q_{\frac{\alpha}{2}}(\{\tilde R_i\}_{i\notin\mathcal{I}_{ca}^l})\big\},\\
    &b_{up}(x_j,\alpha;y_j= l) =expit\big\{L_p(x_j) +q_{1-\frac{\alpha}{2}}(\{\tilde R_i\}_{i\in\mathcal{I}_{ca}^l})\big\},\\
    &b_{up}(x_j,\alpha;y_j\neq l) =expit\big\{L_p(x_j) +q_{1-\frac{\alpha}{2}}(\{\tilde R_i\}_{i\notin\mathcal{I}_{ca}^l})\big\}.
\end{align*}\\
\vspace{-2 mm}
\ENDFOR
\STATE Obtain faith interval for $TPR(\lambda)$ and $FPR(\lambda)$ from
\vspace{-2mm}
{\scriptsize
\begin{align*}
    &CI_{\lambda,l}^{sen}(\mathcal{D}_{tst}^l,\alpha) = \bigg[\frac{1}{|\mathcal{D}_{tst}^l|}\sum_{j\in\mathcal{I}_{tst}^l} {\bf 1}\big(b_{lo}(x_j,\alpha;y_j=l)>\lambda\big),\quad \frac{1}{|\mathcal{I}_{tst}^l|}\sum_{j\in\mathcal{I}_{tst}^l} {\bf 1}\big(b_{up}(x_j,\alpha;y_j=l)>\lambda\big) \bigg]\\    
    &CI_{\lambda,l}^{spe}(\overline{\mathcal{D}_{tst}^l},\alpha) = \bigg[\frac{1}{|\overline{\mathcal{D}_{tst}^l}|}\sum_{j\notin\mathcal{I}_{tst}^l} {\bf 1}\big(b_{lo}(x_j,\alpha;y_j\neq l)>\lambda\big),\quad \frac{1}{|\overline{\mathcal{D}_{tst}^l}|}\sum_{j\notin\mathcal{I}_{tst}^l} {\bf 1}\big(b_{up}(x_j,\alpha;y_j\neq l)>\lambda\big), \bigg]
\end{align*}}\\
\vspace{-2mm}
\RETURN Collect $\big\{ \big(CI_{\lambda,l}^{sen}(\mathcal{D}_{tst}^l,\alpha),CI_{\lambda,l}^{spe}(\overline{\mathcal{D}_{tst}^l},\alpha)\big):\lambda\in[0,1]\big\}$ and plot the ROC curve bands. 
\end{algorithmic}
\end{algorithm}

For the non-iid case, the algorithm is similar. Combining ROCs for different classes is still an open question in the literature, and in practice, people often use simple methods such as taking the average of all of them. However, if there is a rare label with very few data points falling into this category, the ROC can perform poorly due to its small sample size. We suggest taking a weighted average by the size of each labeled group to down-weight the potential unreliable confidence bands from rare labels.

\section{Proof of the theorems}
\subsection{Proof of Proposition 2.1}
Define $G_n(x) = \frac{1}{|\mathcal{D}_{ca}|}\sum_{i\in\mathcal{D}_{ca}} {\bf 1}(R_i\le x)$.
\begin{align}  &Pr\big(\pi_{new}\in\widetilde{PI}^{naive}(x_{new},\alpha)\big) = Pr\bigg( q_{\alpha/2}(\{R_i\}_{i\in\mathcal{I}_{ca}})\le R_{new}\le q_{1-\alpha/2}(\{R_i\}_{i\in\mathcal{I}_{ca}}) \bigg)\nonumber\\
&=Pr\bigg(\frac{|\mathcal{D}_{ca}|+1}{|\mathcal{D}_{ca}|}\alpha/2\le G_n(R_{new})\le \frac{|\mathcal{D}_{ca}|+1}{|\mathcal{D}_{ca}|}(1-\alpha/2)\bigg)\ge 1-\alpha\label{eq:pf2.1}
\end{align}
We know that
\begin{align*}
    &Pr\big(G_n(R_{new})< \frac{|\mathcal{D}_{ca}|+1}{|\mathcal{D}_{ca}|}\alpha/2 \big) \\
    &= Pr\bigg( \frac{1}{|\mathcal{D}_{ca}|}\sum_{i\in\mathcal{D}_{ca}\cup\{new\}} {\bf 1}(R_i\le R_{new}) < \frac{|\mathcal{D}_{ca}|+1}{|\mathcal{D}_{ca}|}\alpha/2+\frac{1}{|\mathcal{D}_{ca}|}\bigg)\\
    &=\mathbb{E} {\bf 1}\bigg(\frac{1}{|\mathcal{D}_{ca}|}\sum_{i\in\mathcal{D}_{ca}\cup\{new\}} {\bf 1}(R_i\le R_{new})<\frac{|\mathcal{D}_{ca}|+1}{|\mathcal{D}_{ca}|}\alpha/2+\frac{1}{|\mathcal{D}_{ca}|} \bigg) \\
    &= \mathbb{E}\bigg(\frac{1}{|\mathcal{D}_{ca}|+1}\sum_{j\in\mathcal{D}_{ca}\cup\{new\}}{\bf 1}\big(\frac{1}{|\mathcal{D}_{ca}|}\sum_{i\in\mathcal{D}_{ca}\cup\{new\}} {\bf 1}(R_i\le R_j)<\frac{|\mathcal{D}_{ca}|+1}{|\mathcal{D}_{ca}|}\alpha/2+\frac{1}{|\mathcal{D}_{ca}|} \big) \bigg)
\end{align*}
Now we order $\{R_i\}_{i\in\mathcal{I}_{ca}}\cup\{R_{new}\}$ by $R_{(1)}<\cdots<R_{(|\mathcal{D}_{ca}|)}$, and we can ignore ties because we assume (A2). Given $\frac{|\mathcal{D}_{ca}|+1}{|\mathcal{D}_{ca}|}\alpha/2+\frac{1}{|\mathcal{D}_{ca}|}$, there exists $k\in\{0,1,\cdots,|\mathcal{D}_{ca}|\}$, such that $k_1<(|\mathcal{D}_{ca}|+1)\alpha/2+1\le k_1+1$, thus
\begin{align*}
    Pr\big(G_n(R_{new})\le \frac{|\mathcal{D}_{ca}|+1}{|\mathcal{D}_{ca}|}\alpha/2 \big) = \frac{k_1}{|\mathcal{D}_{ca}|+1}< \alpha/2+\frac{1}{|\mathcal{D}_{ca}|+1}
\end{align*}
Similarly, we have that for $k_2<(|\mathcal{D}_{ca}|+1)(1-\alpha/2)+1\le k_2+1$
\begin{align*}
    Pr\big(G_n(R_{new})\le \frac{|\mathcal{D}_{ca}|+1}{|\mathcal{D}_{ca}|}(1-\alpha/2) \big) = \frac{k_2+1}{|\mathcal{D}_{ca}|+1}\ge 1-\alpha/2 +\frac{1}{|\mathcal{D}_{ca}|+1}
\end{align*}
So combining both sides, we have~(\ref{eq:pf2.1}).

\vspace{5mm}
\subsection{ Proof of Theorem 2.1}
Define $\tilde G_n(x) = \frac{1}{|\mathcal{D}_{ca}|}\sum_{i\in\mathcal{D}_{ca}} {\bf 1}(\tilde R_i\le x)$. We need to prove
\begin{align*}
    1-\alpha-o(1)&\le Pr\bigg(q_{\alpha/2}(\{\tilde R_i\}_{i\in\mathcal{I}_{ca}})\le R_{new}\le q_{1-\alpha/2}(\{\tilde R_i\}_{i\in\mathcal{I}_{ca}}) \bigg)\\
    &\le Pr\bigg(\frac{|\mathcal{D}_{ca}|+1}{|\mathcal{D}_{ca}|}\alpha/2\le \tilde G_n(R_{new})\le \frac{|\mathcal{D}_{ca}|+1}{|\mathcal{D}_{ca}|}(1-\alpha/2)\bigg)
\end{align*}

First, notice that as $|\mathcal{D}_{ca}|\to\infty $,
\begin{align}
    &|G_n(x)-\tilde G_n(x+\delta_n)| \le \frac{1}{|\mathcal{D}_{ca}|}\sum_{i\in\mathcal{D}_{ca}} \big|{\bf 1}(R_i\le x)-{\bf 1}(\tilde R_i\le x+\delta_n)\big|\nonumber\\
    &= \frac{1}{|\mathcal{D}_{ca}|}\sum_{i\in\mathcal{D}_{ca}} \bigg({\bf 1}(R_i\le x,\tilde R_i>x+\delta_n)+{\bf 1}(R_i>x,\tilde R_i\le x+\delta_n)\bigg)\nonumber\\
    &\le \frac{1}{|\mathcal{D}_{ca}|}\sum_{i\in\mathcal{D}_{ca}} {\bf 1}(|R_i-\tilde R_i|>\delta_n) + \frac{1}{|\mathcal{D}_{ca}|}\sum_{i\in\mathcal{D}_{ca}} {\bf 1}(R_i>x,\tilde R_i\le x+\delta_n)\label{eq:pf2.1_1}
\end{align}
For the first term in ~(\ref{eq:pf2.1_1}), we have that 
$$\lim_{|\mathcal{D}_{ca}|\to\infty} \frac{1}{|\mathcal{D}_{ca}|}\sum_{i\in\mathcal{D}_{ca}} {\bf 1}(|R_i-\tilde R_i|>\delta_n) = \mathbb  P(|R_i-\tilde R_i|>\delta_n)=0$$
For the second term, we have that
\begin{align*}
    &\lim_{|\mathcal{D}_{ca}|\to\infty}\frac{1}{|\mathcal{D}_{ca}|}\sum_{i\in\mathcal{D}_{ca}} {\bf 1}(R_i>x,\tilde R_i\le x+\delta_n) =\lim_{|\mathcal{D}_{ca}|\to\infty}Pr(R_i>x,\tilde R_i\le x+\delta_n) \\
    &= \lim_{|\mathcal{D}_{ca}|\to\infty}Pr(R_i>x,\tilde R_i\le x+\delta_n\big| |R_i-\tilde R_i|>\delta_n )Pr(|R_i-\tilde R_i|>\delta_n) \\
    &\quad +\lim_{|\mathcal{D}_{ca}|\to\infty}Pr(R_i>x,\tilde R_i\le x+\delta \big| |R_i-\tilde R_i|\le\delta_n )Pr(|R_i-\tilde R_i|\le\delta_n)\\
    &\le \lim_{|\mathcal{D}_{ca}|\to\infty}Pr(|R_i-\tilde R_i|>\delta_n) + \lim_{|\mathcal{D}_{ca}|\to\infty}Pr(x\le R_i\le x+2\delta_n)\\
    &\le F_R(x+2\delta_n)-F_R(x)\le 2C\delta_n
\end{align*}

Second, from Proposition 2.1, we know that for any small $\gamma>0$, $Pr\big(G_n(R_{new})\le\gamma\big)\le \gamma +O(|1/\mathcal{D}_{ca}|) $. Thus
\begin{align*}
    &\lim_{|\mathcal{D}_{ca}|\to\infty}Pr\big(\tilde G_n(R_{new})\le \alpha \big) \\
    &= \lim_{|\mathcal{D}_{ca}|\to\infty}Pr\bigg(G_n(R_{new}-\delta_n)\le \alpha+\big(G_n(R_{new}-\delta_n)-\tilde G_n(R_{new})\big) \bigg)\\
    &\le \lim_{|\mathcal{D}_{ca}|\to\infty}Pr\big(G_n(R_{new}-\delta_n)\le \alpha+2C\delta_n\big)\le \alpha +2C\delta_n
\end{align*}
Similarly for the other side of the inequality, thus we proof the theorem.

\vspace{5mm}
\subsection{ Proof of Theorem 3.1}
For the first part,
from assumption (A1), we know that given $y_{new}=k$, $x_{new}$ and  $\{x_i:i\in\mathcal{I}_{ca}^k\}$ will be exchangeable. Then we have that
\begin{align*}
    &\lim_{|\mathcal{D}_{ca}|\to\infty}P\bigg(\pi(x_{new})\in CI^{(k)}(x_{new},\alpha)|y_{new}=k \bigg)\\
    &=\lim_{|\mathcal{D}_{ca}|\to\infty}P\bigg(q_{\alpha/2}(\{\tilde R_i\}_{i\in\mathcal{I}_{ca}^k})\le R_{new} \le q_{1-\alpha/2}(\{\tilde R_i\}_{i\in\mathcal{I}_{ca}^k})|y_{new}=k\bigg)\\
    &\le \alpha + C_k\delta_n^{(k)}
\end{align*}
The last inequality comes directly from Theorem 2.1. 

For the second part, we know that
\begin{align*}
    &P\big(p_n(x_{new})\in CI^{(k)}(x_{new},\alpha)|y_{new}=k \big)\\
    &=P\big( q_{\alpha/2}(\{\tilde R_i\}_{i\in\mathcal{I}_{ca}^k})\le 0\le q_{1-\alpha/2}(\{\tilde R_i\}_{i\in\mathcal{I}_{ca}^k}) |y_{new}=k \big)
\end{align*}
So we only need to show that $ P\big( q_{\alpha/2}(\{\tilde R_i\}_{i\in\mathcal{I}_{ca}^k})< 0\big)\to 1 $ and $ P\big( q_{1-\alpha/2}(\{\tilde R_i\}_{i\in\mathcal{I}_{ca}^k})> 0\big)\to 1 $. as $|\mathcal{D}_{ca}^k|,|\mathcal{D}_{tr}|\to\infty$.

From the proof of Theorem 2.1, we know that $|G_n(x)-\tilde G_n(x+\delta_n)|\le O\big(\epsilon_n,\delta_n,\frac{1}{|\mathcal{D}_{ca}|}\big)$.
We can do the same thing for $\mathcal{D}_{ca}^k$ and have 
\begin{equation*}
    \big|q_{\alpha/2}(\{\tilde R_i\}_{i\in\mathcal{I}_{ca}^k}) - q_{\alpha/2}(\{ R_i\}_{i\in\mathcal{I}_{ca}^k})\big| = \big|o_p(1)\big|
\end{equation*}

With (A2'), we know that the density function of $F_{R,k}$ exists and denote it as $f_{R,k}$. With (A1), we know that given $\mathcal{D}_{tr}$, $\{R_i\}_{i\in\mathcal{I}_{ca}}$ are iid continuous random variables, so as $|\mathcal{D}_{ca}|\to\infty$,
\begin{align*}
    \sqrt{|\mathcal{D}_{ca}|}\big(q_{\alpha/2}(\{ R_j^k\}_{j\in\mathcal{I}_{ca}}) - F_{R,k}^{-1}(\alpha/2) \big)\to N\bigg(0,\frac{\alpha/2(1-\alpha/2)}{f_{R,k}\big(F_{R,k}^{-1}(\alpha/2)\big)^2}\bigg)
\end{align*}
Thus from Slutsky's theorem, we have $\big|q_{\alpha/2}(\{\tilde R_i\}_{i\in\mathcal{I}_{ca}^k}) - F_{R,k}^{-1}(\alpha/2)\big| = \big|o_p(1)\big| $.
Similarly, we have $\big|q_{1-\alpha/2}(\{\tilde R_i\}_{i\in\mathcal{I}_{ca}^k}) - F_{R,k}^{-1}(1-\alpha/2)\big| = \big|o_p(1)\big| $ for the other side of the inequality, and the theorem is proved.

\vspace{5mm}
\subsection{ Proof of Theorem 3.2}
For the first part of the theorem, we only need to show the coverage rate for $TPR^{(o)}(\lambda)$ and the coverage rate for $FPR^{(o)}(\lambda)$ is exactly the same.
\begin{align}
    &P\big(TPR^{(o)}(\lambda) \in CI_{\lambda}^{sen}(\mathcal{D}_{tst}^1,\alpha)\big) = \nonumber\\
    &P\bigg\{
    \frac{1}{|\mathcal{I}_{tst}^1|}\sum_{j \in \mathcal{I}_{tst}^1} \mathbf{1}\big[b_{lo}(x_j,\alpha;1) > \lambda\big] \leq \frac{1}{|\mathcal{I}_{tst}^1|}\sum_{j \in \mathcal{I}_{tst}^1} \mathbf{1}\big[\pi(x_j) > \lambda\big]\text{, and }\label{eq:pf3.2_1}  \\
    &\quad \quad\quad
    \frac{1}{|\mathcal{I}_{tst}^1|}\sum_{j \in \mathcal{I}_{tst}^1} \mathbf{1}\big[\pi(x_j) > \lambda\big] \leq \frac{1}{|\mathcal{I}_{tst}^1|}\sum_{j \in \mathcal{I}_{tst}^1} \mathbf{1}\big[b_{up}(x_j,\alpha;1) > \lambda\big]\bigg\} \nonumber
\end{align}
Denote test data's conformity score $R_j = L_{\pi}(x_j)-L_p(x_j)$ for $j\in\mathcal{I}_{tst}$. For the left side
\begin{align}
    & P\bigg\{ \frac{1}{|\mathcal{I}_{tst}^1|}\sum_{j\in\mathcal{I}_{tst}^1} {\bf 1}\big[\pi(x_{j})>\lambda\big] < \frac{1}{|\mathcal{I}_{tst}^1|}\sum_{j\in\mathcal{I}_{tst}^1} {\bf 1}\big[b_{lo}(x_j,\alpha;1)>\lambda\big]\bigg\}\nonumber\\
    & = P\bigg\{ \sum_{j\in\mathcal{I}_{tst}^1} {\bf 1}\big[\pi(x_{j})>\lambda\big] - {\bf 1}\big[b_{lo}(x_j,\alpha;1)>\lambda\big]< 0\bigg\}\nonumber\\
    & = P\bigg\{ \sum_{j\in\mathcal{I}_{tst}^1} {\bf 1}\big[\pi(x_{j})>\lambda\big] - {\bf 1}\big[ expit\big\{L_p(x_j) +q_{\alpha/2}(\{\tilde R_i\}_{i\in\mathcal{I}_{ca}^1})\big\} >\lambda\big]< 0\bigg\}\nonumber\\
    & = P\bigg\{ \sum_{j\in\mathcal{I}_{tst}^1} {\bf 1}\big[L_{\pi}(x_j)>logit\big(\lambda\big)\big] < \\
    &\quad  \quad \; \; \; \, \sum_{j\in\mathcal{I}_{tst}^1}{\bf 1}\big[ L_{\pi}(x_j)  >logit(\lambda)+R_j- q_{\alpha/2}(\{\tilde R_i\}_{i\in\mathcal{I}_{ca}^1})\big]\bigg\} \nonumber
\end{align}
for the right hand side inside the last term,
\begin{align*}
   &\sum_{j\in\mathcal{I}_{tst}^1}{\bf 1}\big[ L_{\pi}(x_j)  >logit(\lambda)+R_j- q_{\alpha/2}(\{\tilde R_i\}_{i\in\mathcal{I}_{ca}^1})\big]\\
   & = \sum_{j\in\mathcal{I}_{tst}^1}{\bf 1}\big[ L_{\pi}(x_j)   >logit(\lambda)+R_j- q_{\alpha/2}(\{\tilde R_i\}_{i\in\mathcal{I}_{ca}^1}) \big]{\bf 1}\big[L_{\pi}(x_j) \le logit\big(\lambda\big)\big]\\
   &\quad\quad\quad+ {\bf 1}\big[ L_{\pi}(x_j)   >logit(\lambda)+R_j- q_{\alpha/2}(\{\tilde R_i\}_{i\in\mathcal{I}_{ca}^1}) \big]{\bf 1}\big[L_{\pi}(x_j) >logit\big(\lambda\big)\big] \\
   &\le \sum_{j\in\mathcal{I}_{tst}^1} {\bf 1}\big[R_j< q_{\alpha/2}(\{\tilde R_i\}_{i\in\mathcal{I}_{ca}^1}) \big]{\bf 1}\big[L_{\pi}(x_j) \le logit\big(\lambda\big)\big] \\
   &\quad\quad\quad+ {\bf 1}\big[ L_{\pi}(x_j)   >logit(\lambda)+R_j- q_{\alpha/2}(\{\tilde R_i\}_{i\in\mathcal{I}_{ca}^1}) \big]{\bf 1}\big[L_{\pi}(x_j) >logit\big(\lambda\big)\big]\\
   &\le \sum_{j\in\mathcal{I}_{tst}^1}{\bf 1}\big[R_j< q_{\alpha/2}\big(\{ \tilde R_i\}_{i\in\mathcal{I}_{ca}^1}\big) \big] \\
   &\quad\quad\quad+ \sqrt{\sum_{j\in\mathcal{I}_{tst}^1} {\bf 1}\big[ L_{\pi}(x_j)  >logit(\lambda)+R_j- q_{\alpha/2}(\{ \tilde R_i\}_{i\in\mathcal{I}_{ca}^1}) \big] \sum_{j\in\mathcal{I}_{tst}^1}{\bf 1}\big[L_{\pi}(x_j)>logit\big(\lambda\big)\big]}
\end{align*}
The last inequality comes from Cauchy inequality and also the fact that ${\bf 1}^2(A) = {\bf 1}(A)$ for any event $A$. 
From the previous proof, we have $|q_{\alpha/2}(\{\tilde R_i\}_{i\in\mathcal{I}_{ca}^k}) - q_{\alpha/2}(\{ R_i\}_{i\in\mathcal{I}_{ca}^k})| = o_p(1)$ and $|q_{\alpha/2}(\{ R_i\}_{i\in\mathcal{I}_{ca}^k}) - q_{\alpha/2}(\{ R_i\}_{i\in\mathcal{I}_{tst}^k})|=o_p(1) $, thus $|q_{\alpha/2}(\{\tilde R_i\}_{i\in\mathcal{I}_{ca}^1}) - q_{\alpha/2}(\{ R_i\}_{i\in\mathcal{I}_{tst}^1})|=o_p(1)$. Now plug in $\lambda = \pi(x_s)$ and we have that

\begin{align}
    &P\bigg\{ \frac{1}{|\mathcal{I}_{tst}^1|}\sum_{j\in\mathcal{I}_{tst}^1} {\bf 1}\big[\pi(x_{j})>\lambda\big]< \frac{1}{|\mathcal{I}_{tst}^1|}\sum_{j\in\mathcal{I}_{tst}^1} {\bf 1}\big[b_{lo}(x_j,\alpha;1)>\lambda\big]\bigg\}\nonumber\\
    &\le P\bigg\{ \sum_{j\in\mathcal{I}_{tst}^1} {\bf 1}\big[L_{\pi}(x_j)>L_{\pi}(x_s)\big]< \sum_{j\in\mathcal{I}_{tst}^1}{\bf 1}\bigg[R_j< q_{\alpha/2}\big(\{ R_i\}_{i\in\mathcal{I}_{tst}^1}\big)+|o_p(1)| \bigg] + \nonumber\\
    &\quad \sqrt{\sum_{j\in\mathcal{I}_{tst}^1} {\bf 1}\bigg[ L_{\pi}(x_j)  >L_{\pi}(x_s)+R_j- q_{\alpha/2}(\{ R_i\}_{i\in\mathcal{I}_{tst}^1})-|o_p(1)| \bigg] \sum_{j\in\mathcal{I}_{tst}^1}{\bf 1}\big[L_{\pi}(x_j)>L_{\pi}(x_s)\big]}\bigg\} \nonumber\\
   &= \mathbb{E}\frac{1}{|\mathcal{I}_{tst}^1|}\sum_{s\in\mathcal{I}_{tst}^1}{\bf 1}\bigg\{\sum_{j\in\mathcal{I}_{tst}^1} {\bf 1}\big[L_{\pi}(x_j)>L_{\pi}(x_s)\big]< \sum_{j\in\mathcal{I}_{tst}^1}{\bf 1}\bigg[R_j< q_{\alpha/2}\big(\{ R_i\}_{i\in\mathcal{I}_{tst}^1}\big)+|o_p(1)| \bigg] + \nonumber\\
    &\quad \sqrt{\sum_{j\in\mathcal{I}_{tst}^1} {\bf 1}\bigg[ L_{\pi}(x_j)  >L_{\pi}(x_s)+R_j- q_{\alpha/2}(\{ R_i\}_{i\in\mathcal{I}_{tst}^1})-|o_p(1)| \bigg] \sum_{j\in\mathcal{I}_{tst}^1}{\bf 1}\big[L_{\pi}(x_j)>L_{\pi}(x_s)\big]}\bigg\} \nonumber\\
   &\le \frac{1}{|\mathcal{I}_{tst}^1|}\bigg\{\mathbb{E}\sum_{s:q_s\le\alpha/2} 1 + \mathbb{E}\sum_{s:q_s>\alpha/2} {\bf 1}\big[q_s\le \alpha/2+\sqrt{{\bf 1}(q_s\le \alpha/2)q_s}  \big]\bigg\} = \alpha/2\label{eq:pf3.2_2}
\end{align}
Here $q_s$ denotes the quantile of $logit\big(\pi(x_s)\big)$ among $\{logit\big(\pi(x_i)\big)\}_{i\in\mathcal{I}_{tst}^1}$, and (\ref{eq:pf3.2_2}) comes from the fact that quantile functions are convex, thus 
\begin{align*}
    q_{\alpha/2}\big(\{L_p(x_i)\}_{i\in\mathcal{I}_{tst}^1} \big) + q_{\alpha/2}\big(\{R_i\}_{i\in\mathcal{I}_{tst}^1} \big)\ge q_{\alpha/2}\big(\{L_{p}(x_i)+R_i\}_{i\in\mathcal{I}_{tst}^1}\big) =   q_{\alpha/2}\big(\{L_{\pi}(x_i)\}_{i\in\mathcal{I}_{tst}^1}\big)
\end{align*}


The other side of the inequality of~(\ref{eq:pf3.2_1}) is similar and we have 
\begin{equation*}
    P\bigg( \frac{1}{|\mathcal{I}_{tst}^1|}\sum_{j\in\mathcal{I}_{tst}^1} {\bf 1}\big(\pi(x_{j})>\lambda\big)\le \frac{1}{|\mathcal{I}_{tst}^1|}\sum_{j\in\mathcal{I}_{tst}^1} {\bf 1}\big(c_{j,u}^{(1)}(\alpha)>\lambda\big)\bigg)\le 1-\alpha/2
\end{equation*}
Thus prove the first part of Theorem 3.2. For the second part, we only need to prove one side of the inequality in the following:
\begin{align*}
    &\lim_{|\mathcal{D}_{tst}|,|\mathcal{D}_{ca}^1|,|\mathcal{D}_{tr}|\to\infty} TPR(\lambda)-\frac{1}{|\mathcal{I}_{tst}^1|}\sum_{j\in\mathcal{I}_{tst}^1}{\bf 1}\big(b_{up}(x_j,\alpha;1)>\lambda\big)\\
    &= \lim_{|\mathcal{D}_{tst}|,|\mathcal{D}_{ca}^1|,|\mathcal{D}_{tr}|\to\infty}\frac{1}{|\mathcal{I}_{tst}^1|} \sum_{j\in\mathcal{I}_{tst}^1} \bigg({\bf 1}\big(p_n(x_j)>\lambda\big) - {\bf 1}\big(b_{up}(x_j,\alpha;1)>\lambda\big)\bigg)\\
    &\le \lim_{|\mathcal{D}_{tst}|,|\mathcal{D}_{ca}^1|,|\mathcal{D}_{tr}|\to\infty}\frac{1}{|\mathcal{I}_{tst}^1|}\sum_{j\in\mathcal{I}_{tst}^1}{\bf 1}\big(p_n(x_j)>\lambda,b_{up}(x_j,\alpha;1)\le \lambda\big)\\
    &\le \lim_{|\mathcal{D}_{tst}|,|\mathcal{D}_{ca}^1|,|\mathcal{D}_{tr}|\to\infty}\frac{1}{|\mathcal{I}_{tst}^1|}\sum_{j\in\mathcal{I}_{tst}^1}{\bf 1}\big(p_n(x_j)>b_{up}(x_j,\alpha;1)\big)\\
    &=\lim_{|\mathcal{D}_{ca}^1|,|\mathcal{D}_{tr}|\to\infty} Pr\big(p_n(x_j)>c_{j,u}^{(1)}|y_j=1\big)= 0
\end{align*}
Similarly, we have the other side of the inequality, thus proof the second part.

\section{Simulation studies}
In this section, we present two simulation studies to validate our theoretical results and compare the ROC curve bands for different classification algorithms. In the first study, we investigate the impact of data size on the coverage probability for different classification models and provide a numerical example to illustrate the dataset size required to ensure the correct level of uncertainty. In the second study, we compare the performance of classification models with different levels of biases in terms of the coverage and average length of their confidence intervals. The differences can also be assessed by examining the ROC curve band width.

For both studies, we report the coverage probability of the intervals \( CI(x_{new}, \alpha; \) \(y_{new} = k) \) and \( CI_c(x_{new}, \alpha; y_{new} = k) \), \( k = 1, 0 \), as developed in Section~\ref{sec: methodology}. According to the theoretical results, we are interested in two types of coverage for both intervals: (a) whether the interval will cover the true (unobserved) \(\pi(\cdot)\), and (b) whether the interval will cover the predicted (observed) \( p_n(\cdot) \). We use the following empirical coverage to estimate these two types of coverage:
\begin{align*}
    &\text{True \(\pi(x)\) coverage: } \widehat{Cov}_{\pi}^{(k)} = \frac{1}{|\mathcal{D}_{tst}^k|}\sum_{j \in \mathcal{I}_{tst}^k} \mathbf{1}\big[\pi(x_j) \in CI(x_{new}, \alpha; y_{new} = k) \big] \\
    &\text{Predicted \( p_n(x) \) coverage: } \widehat{Cov}_{p}^{(k)} = \frac{1}{|\mathcal{D}_{tst}^k|}\sum_{j \in \mathcal{I}_{tst}^k} \mathbf{1}\big[p_n(x_j) \in CI(x_{new}, \alpha; y_{new} = k) \big]
\end{align*}
Similarly, we define the two types of empirical coverages for \( CI_c(x_{new}, \alpha; y_{new} = k) \).

For the rest of this section, we introduce our two simulation studies in detail and show the results that validate the theoretical findings developed in the previous sections.

\subsection{Simulation Study 1: Sample Size}

In this study, we will focus on the impact of sample size on the confidence interval and ROC curve bands for classification algorithms trained on different models.

Consider generating observed data from the following model:
\begin{equation}\label{eq: sim1 generate data}
    y_i \sim \text{Ber}(\pi(x_i)), \text{ with }
    \text{logit}(\pi(x_i)) = \beta_0 + \beta_1 x_{i1} + \beta_2 x_{i2} + \beta_3 x_{i3}, \quad i = 1, 2, \cdots, n.
\end{equation}

Here, \( x_i = (x_{i1}, x_{i2}, x_{i3})^T \in \mathbb{R}^3 \) are the vector of covariates and are generated from \( N((0, 0, 0)^T, \Sigma) \). We set the covariance matrix \(\Sigma\) to be \(\Sigma_{ii} = 1\), \(\Sigma_{ij} = -(-0.1)^{|i-j|}\), \( i \neq j \in \{1, 2, 3\} \). The sample size is chosen from \( n \in [100,2000] \).

The test data are generated from the same model~(\ref{eq: sim1 generate data}), and we consider two cases where the covariates are iid or non-iid with the observed data set. For iid covariates, the \( x_i \)'s of the test data are generated from the same distribution as the observed data. For non-iid covariates, we generate \( x_i \sim N((2, 1, -0.6)^T, \Sigma') \), where \(\Sigma'_{ii} = 0.6 - 0.1i\), \(\Sigma'_{ij} = (0.1)^{|i-j|}\), for \( i \neq j \in \{1, 2, 3\} \).

For the classification algorithm, we assume the data generating mechanism is unknown and use the following three logistic models to construct classification rules:
\begin{itemize}
    \item[-] M1 (correct model specification): \(\text{logit}(\pi(x_i)) = \beta_0' + \beta_1' x_{i1} + \beta_2' x_{i2} + \beta_3' x_{i3}\), \( i = 1, \cdots, n \).
    \item[-] M2 (missing one covariate): \(\text{logit}(\pi(x_i)) = \beta_0' + \beta_2' x_{i2} + \beta_3' x_{i3}\), \( i = 1, \cdots, n \).
    \item[-] M3 (missing two covariates): \(\text{logit}(\pi(x_i)) = \beta_0' + \beta_3' x_{i3}\), \( i = 1, \cdots, n \).
\end{itemize}
We will use MLE to fit the parameters using the R function \texttt{glm}. In the non-iid case, we choose the oracle distance function \(\text{dist}(x_i, x_j) = \left| (\beta_1, \beta_2, \beta_3) (x_i - x_j) \right|\) to construct the neighborhood, with the neighbor size chosen to be 100, so that assumption (A4) will most likely hold.

\begin{figure}[H]
    \centering
    \includegraphics[width=13 cm]{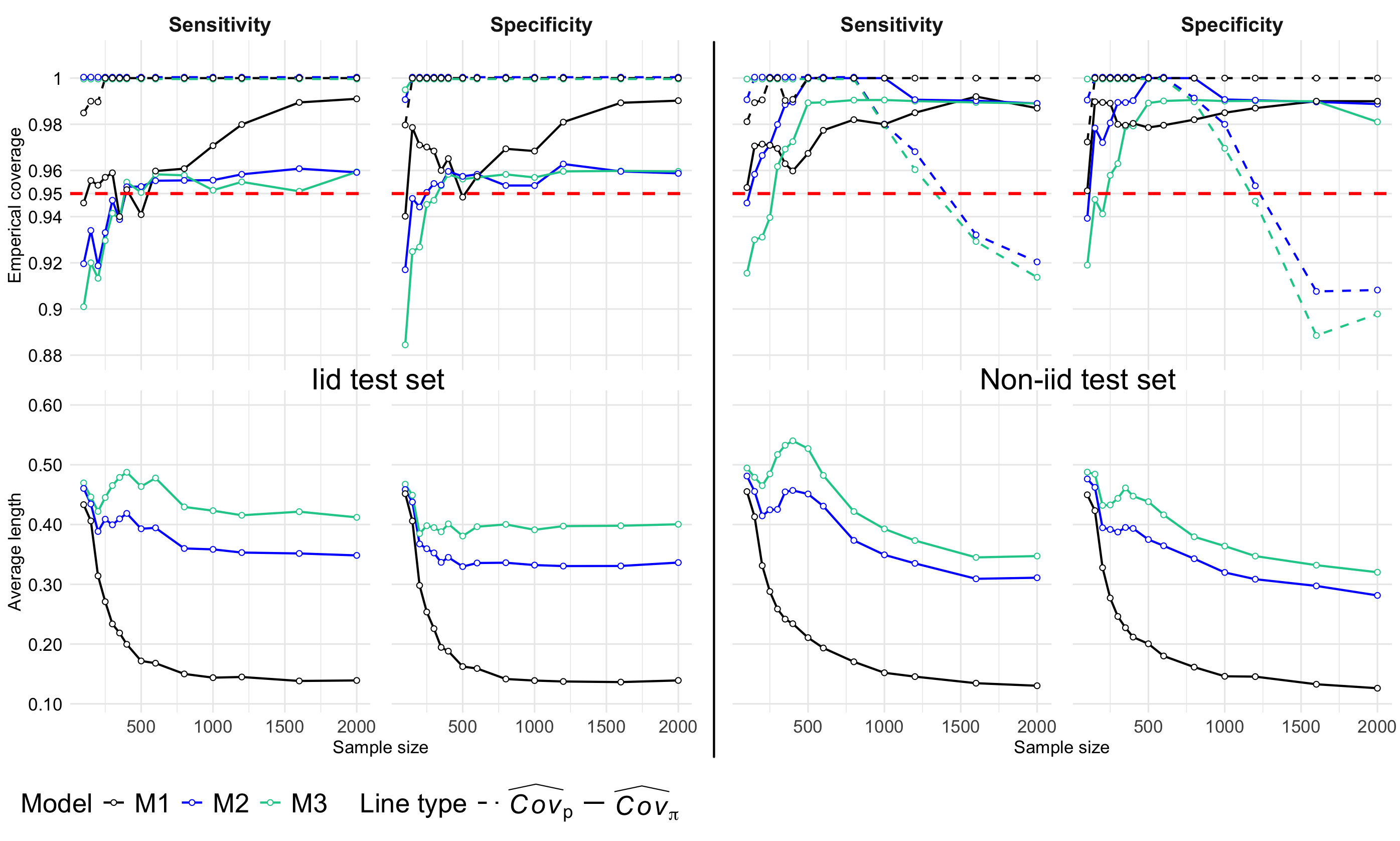}
    \caption{Performances of the Individualized confidence intervals for (M1)-(M3) with different sample size.}
    \label{fig: simulation1_coverage}
\end{figure}

The performance of the confidence intervals is shown in Figure~\ref{fig: simulation1_coverage}. For the iid test data, when the sample size increases, \(\widehat{Cov}_{\pi}\) for all three classification models increases above the target 0.95, and \(\widehat{Cov}_{p}\) quickly converges to 1 as expected. The length first decreases and then becomes stable when the sample size reaches 1000. Comparing the three classification models, the model with the least bias (M1) has the highest coverage with the shortest confidence intervals, while the model with the largest bias (M3) has the longest confidence intervals, indicating that the length of the confidence intervals can reflect the uncertainty level of the classification algorithm. For non-iid test data, all other messages are the same except that the coverage \(\widehat{Cov}_{p}\) is no longer guaranteed for models with large biases (i.e., M2 and M3). This is because the assumption in Theorem~\ref{eq: individual random uncertainty} that \( F_{R,k}^{-1}(\alpha/2) < 0 \) and \( F_{R,k}^{-1}(1 - \alpha/2) > 0 \) no longer holds. In practice, this can be fixed by adding a shift to the \( p_n(\cdot) \) according to the prediction residuals.

\begin{figure}[H]
    \centering
    \includegraphics[width=12 cm]{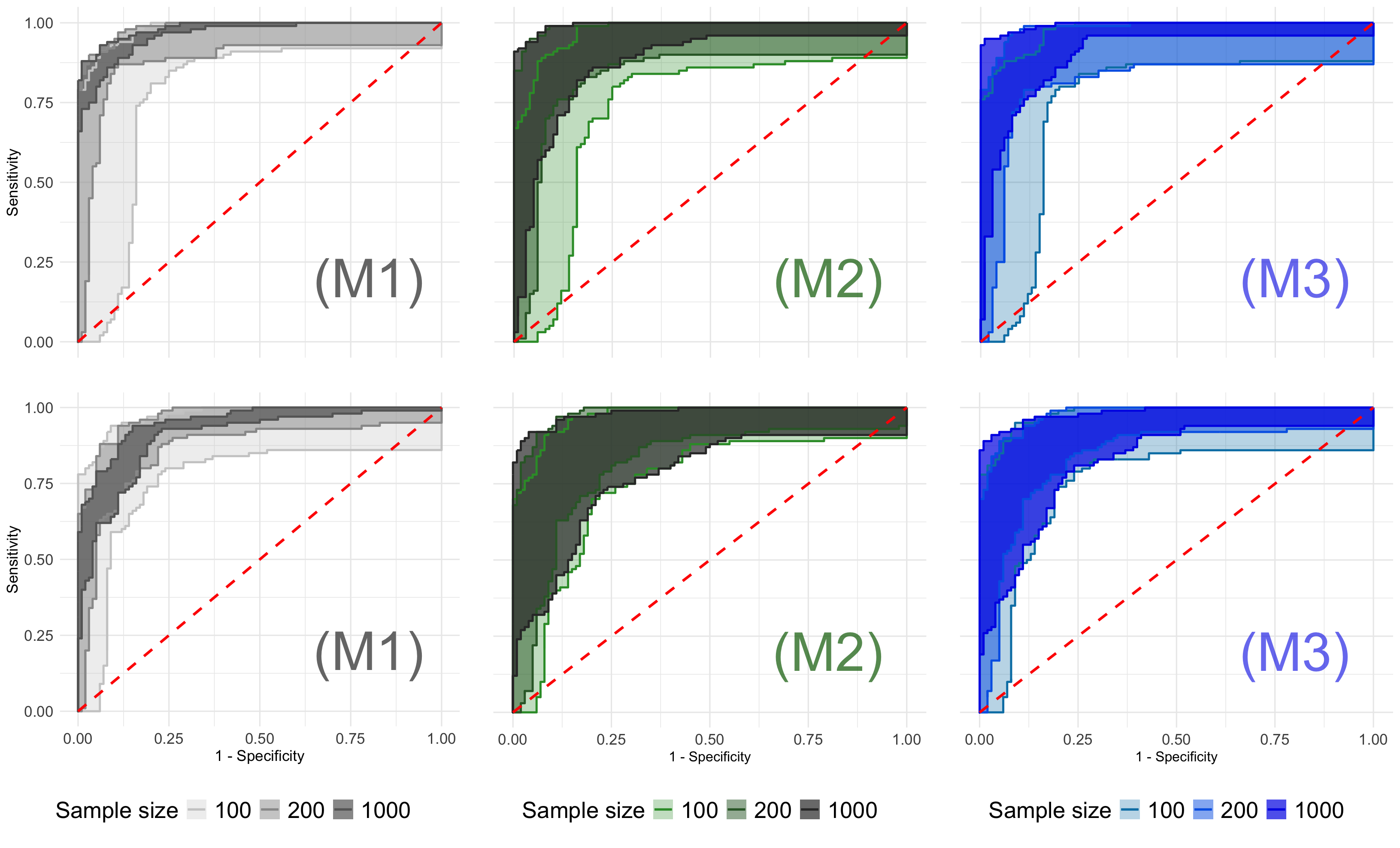}
    \caption{ROC confidence band with fixed specificity for (M1)-(M3) with different sample size. The first row is for iid test data, and the second row is for non-iid test data. The plot for the fixed sensitivity is similar, thus omitted in this section.}
    \label{fig: simulation1_roc}
\end{figure}

To show the performance of the ROC confidence band, we randomly choose the training and test data set in our simulation that is among the average, and show the ROC with confidence bands of the three classification models in Figure~\ref{fig: simulation1_roc}. The messages are clear: the classification model with smaller bias and larger observed data size will have a narrower ROC confidence band, thus smaller uncertainty with respect to both sensitivity and specificity of the test data.


\subsection{Simulation Study 2: Classification model bias}
In this study, we focus on the level of bias in the prediction model and see how it impacts the performance of individual confidence intervals and ROC confidence bands. In practice, when using a logistic model, common sources of bias include (a) missing important predictors and (b) missing interaction terms, etc. So, in this study, we will use two settings to simulate the bias and show its impact on ROC confidence bands' performances.

\subsubsection{Setting 1: missing important predictors}
Consider generating observed data from the following model:
\begin{equation}\label{eq: sim2 generate data}
    y_i \sim \text{Ber}(\pi(x_i)), \text{ with }
    \text{logit}(\pi(x_i)) = \beta_0 + \beta_1 x_{i1} + \beta_2 x_{i2} + \beta_3 x_{i3}, \quad i = 1, 2, \cdots, n.
\end{equation}

Here, \( x_i = (x_{i1}, x_{i2}, x_{i3})^T \in \mathbb{R}^3 \) are the vector of covariates and are generated from \( N((0, 0, 0)^T, \Sigma) \). We set the covariance matrix \(\Sigma\) to be \(\Sigma_{ii} = 1\), \( i = 1, 2, 3 \), \(\Sigma_{12} = \Sigma_{21} = 0.1\) and \(\Sigma_{i3} = \Sigma_{3i} = 0\), \( i = 1, 2 \). The sample size \( n = 500 \). Let \(\beta_1 = \beta_2 = 1\), \(\beta_0 = 1\) and we vary \(\beta_3 \in [0.1, 5.5] \).

The test data are generated from the same model~(\ref{eq: sim2 generate data}), and same as in Study 1, we consider both iid and non-iid cases: for the iid case, the \( x_i \)'s of the test data are generated from the same distribution as the observed data. For the non-iid case, we generate \( x_i \sim N((2, 1, -0.6)^T, \Sigma') \), where \(\Sigma'_{ii} = 0.6 - 0.1i\), \(\Sigma'_{ij} = (0.1)^{|i-j|}\), for \( i \neq j \in \{1, 2, 3\} \). The choice of parameters guarantees that the simulated data are balanced and \( x_{\cdot 3} \) is independent of the other two covariates.

For the classification models, we assume the data generating mechanism is unknown and use the following two logistic models to construct classification rules:
\begin{itemize}
    \item[-] M1 (only use \( x_{\cdot 3} \)): \(\text{logit}(\pi(x_i)) = \beta_0' + \beta_3' x_{i3}\), \( i = 1, \cdots, n \).
    \item[-] M2 (missing \( x_{\cdot 3} \)): \(\text{logit}(\pi(x_i)) = \beta_0' + \beta_1' x_{i1} + \beta_2' x_{i2}\), \( i = 1, \cdots, n \).
\end{itemize}
We will use MLE to fit the parameters using the R function \texttt{glm}. In the non-iid case, we choose the same distance function and neighbor size as in Study 1.

\begin{figure}[H]
    \centering
    \includegraphics[width=12 cm]{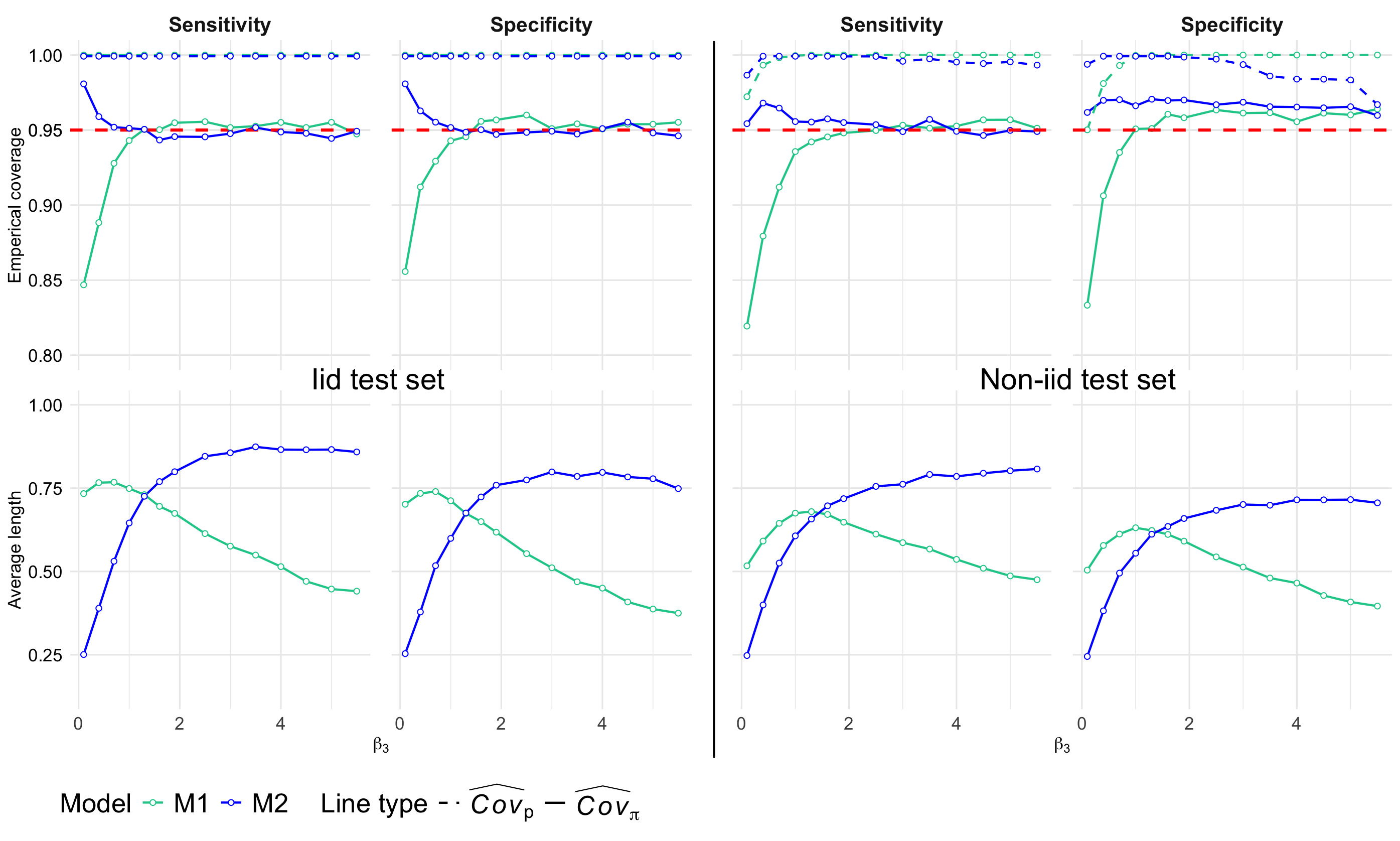}
    \caption{Performances of the Individualized confidence intervals for (M1)-(M3) with different sample size.}
    \label{fig: sim2_coverage}
\end{figure}

The performance of the confidence intervals is shown in Figure~\ref{fig: sim2_coverage}. First, notice that as \(\beta_3\) increases, the bias of M1 decreases and the bias of M2 increases. So, for the iid test data, when the classification model bias decreases, \(\widehat{Cov}_{\pi}\) for both classification models converges to the target 0.95 and \(\widehat{Cov}_{p}\) stays at 1 as expected, and the length decreases, indicating the uncertainty becomes smaller. For non-iid test data, the messages are the same except that when model bias increases, the coverage \(\widehat{Cov}_{p}\) might drop if the assumption in Theorem~\ref{eq: individual random uncertainty} that \( F_{R,k}^{-1}(\alpha/2) < 0 \) and \( F_{R,k}^{-1}(1 - \alpha/2) > 0 \) no longer holds.

\begin{figure}[H]
    \centering
    \includegraphics[width=12 cm]{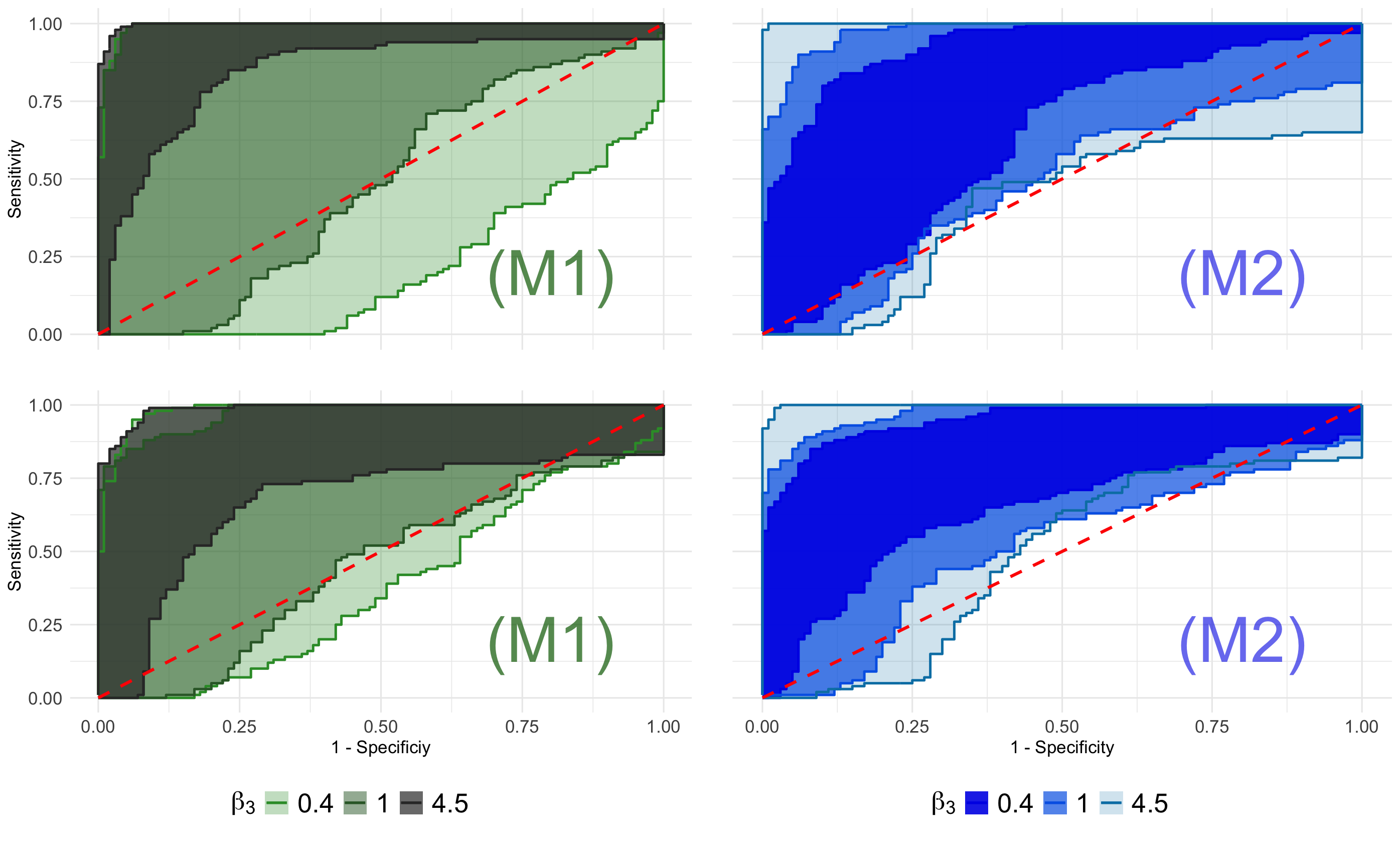}
    \caption{ROC confidence band with fixed specificity for (M1)-(M2) with different sample size. The first row is for iid test data, and the second row is for non-iid test data. The plot for the fixed sensitivity is similar, thus omitted in this section.}
    \label{fig: sim2_roc}
\end{figure}

To show the performance of the ROC confidence band, we randomly choose the training and test data set in our simulation that is among the average, and show the ROC with confidence bands of the two classification models in Figure~\ref{fig: sim2_roc}. The messages are clear: the classification model with smaller bias will have a narrower ROC confidence band, thus smaller uncertainty with respect to both sensitivity and specificity of the test data.


\subsubsection{Setting 2: missing interaction term}

Consider generating observed data from the following model:
\begin{align}
    &y_i \sim \text{Ber}(\pi(x_i)) \nonumber \\
    \text{logit}(\pi(x_i)) = \beta_0 + \beta_1 x_{i1} &+ \beta_2 x_{i2} + \beta_3 x_{i3} + \gamma x_{i1} x_{i3}, \quad i = 1, 2, \cdots, n. \label{eq: sim22 generate data}
\end{align}

Here, \( x_i = (x_{i1}, x_{i2}, x_{i3})^T \in \mathbb{R}^3 \) are the vector of covariates and are generated from \( N((0, 0, 0)^T, \Sigma) \). We set the covariance matrix \(\Sigma\) to be \(\Sigma_{ii} = 1\), \( i = 1, 2, 3 \), \(\Sigma_{12} = \Sigma_{21} = 0.1\), and \(\Sigma_{i3} = \Sigma_{3i} = 0\), \( i = 1, 2 \). The sample size \( n = 500 \). Let \(\beta_1 = \beta_2 = \beta_3 = 1\), \(\beta_0 = 1\) and we vary \(\gamma \in [0.1, 5.5] \).

The test data are generated from the same model~(\ref{eq: sim22 generate data}), and as in Study 1, we consider both iid and non-iid cases: for the iid case, the \( x_i \)'s of the test data are generated from the same distribution as the observed data. For the non-iid case, we generate \( x_i \sim N((1, 1, 2\gamma - 2)^T, \Sigma') \), where \(\Sigma'_{ii} = 0.6 - 0.1i\), \( i = 1, 2, 3 \), \(\Sigma'_{12} = \Sigma'_{21} = (0.1)^{|i-j|} \), and \(\Sigma'_{i3} = \Sigma'_{3i} = 0 \) for \( i = 1, 2 \). The choice of parameters guarantees that the simulated data are balanced and \( x_{\cdot 3} \) is independent of the other two covariates.

For the classification models, we assume the data generating mechanism is unknown and use the following two logistic models to construct classification rules:
\begin{itemize}
    \item[-] M1 (missing interaction term): \(\text{logit}(\pi(x_i)) = \beta_0' + \beta_1' x_{i1} + \beta_2' x_{i2} + \beta_3' x_{i3}\), \( i = 1, \cdots, n \).
    \item[-] M2 (missing \( x_{\cdot 2} \)): \(\text{logit}(\pi(x_i)) = \beta_0' + \beta_1' x_{i1} + \beta_3' x_{i3} + \gamma' x_{i1} x_{i3}\), \( i = 1, \cdots, n \).
\end{itemize}
We will use MLE to fit the parameters using the R function \texttt{glm}. In the non-iid case, we choose the same distance function and neighbor size as in Study 1.

\begin{figure}[H]
    \centering
    \includegraphics[width=12 cm]{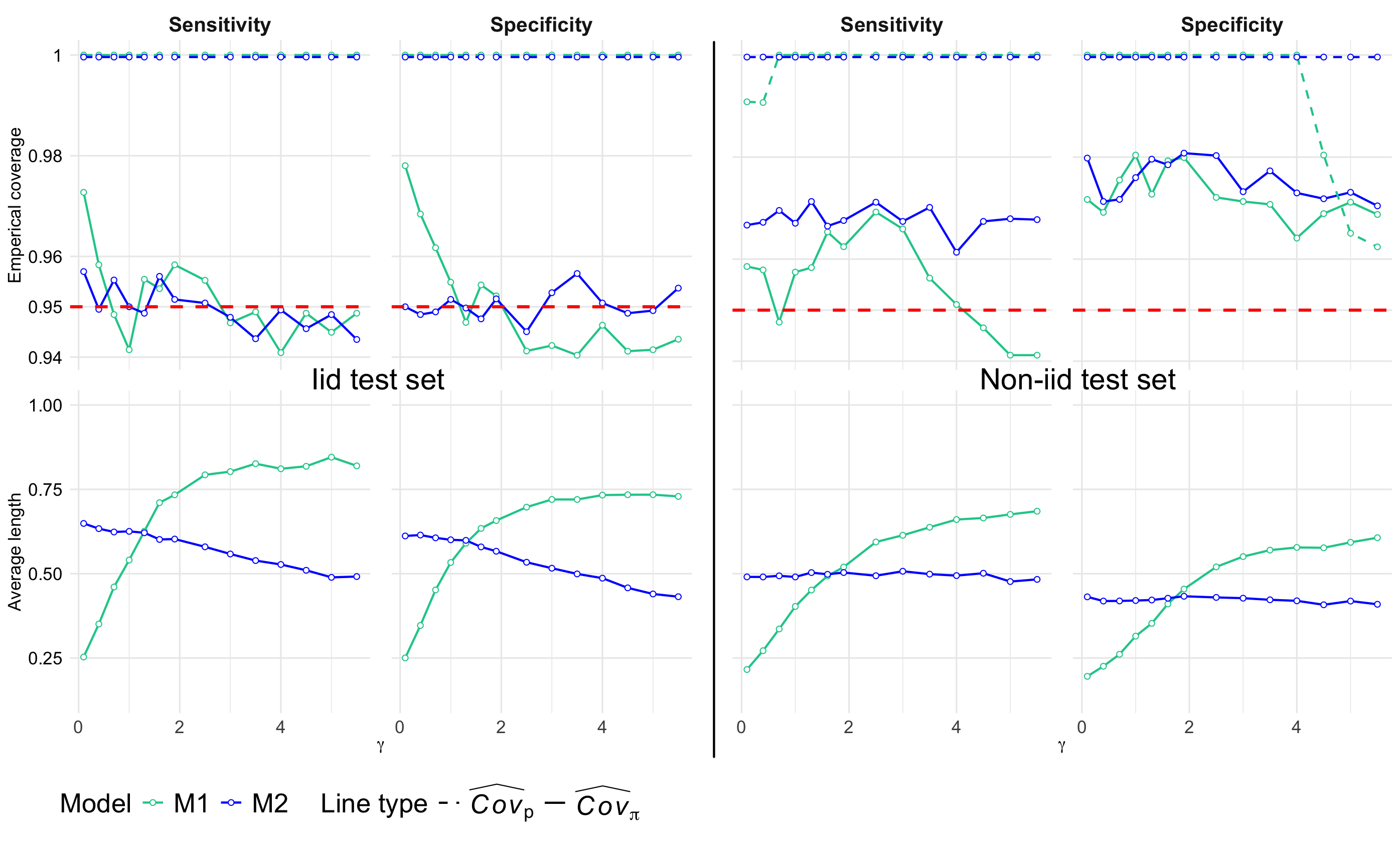}
    \caption{Performances of the Individualized confidence intervals for (M1)-(M3) with different sample size.}
    \label{fig: sim22_coverage}
\end{figure}

The performance of the confidence intervals is shown in Figure~\ref{fig: sim22_coverage}. First, notice that as \(\gamma\) increases, the bias of M1 increases and the bias of M2 decreases. So, for the iid test data, when the classification model bias decreases, \(\widehat{Cov}_{\pi}\) for both classification models converges to the target 0.95 and \(\widehat{Cov}_{p}\) stays at 1 as expected, and the length decreases, indicating the uncertainty becomes smaller. For non-iid test data, the messages are the same except that when model bias increases, the coverage \(\widehat{Cov}_{p}\) might drop if the assumption in Theorem~\ref{eq: individual random uncertainty} that \( F_{R,k}^{-1}(\alpha/2) < 0 \) and \( F_{R,k}^{-1}(1 - \alpha/2) > 0 \) no longer holds.

\begin{figure}[H]
    \centering
    \includegraphics[width=12 cm]{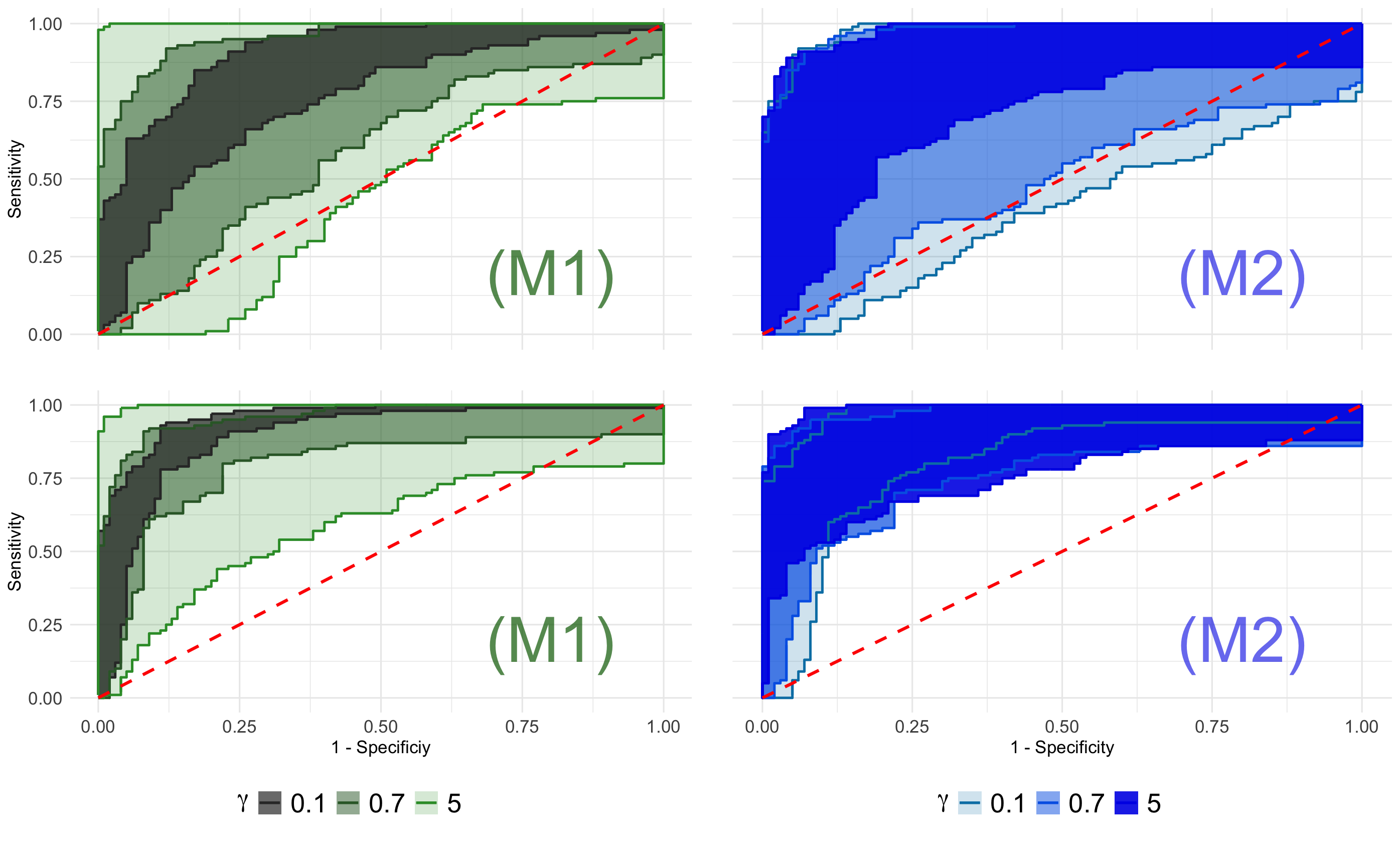}
    \caption{ROC confidence band with fixed specificity for (M1)-(M2) with different sample size. The first row is for iid test data, and the second row is for non-iid test data. The plot for the fixed sensitivity is similar, thus omitted in this section.}
    \label{fig: sim22_roc}
\end{figure}

To show the performance of the ROC confidence band, we randomly choose the training and test data set in our simulation that is among the average, and show the ROC with confidence bands of the two classification models in Figure~\ref{fig: sim22_roc}. The messages are clear: the classification model with smaller bias will have a narrower ROC confidence band, thus smaller uncertainty with respect to both sensitivity and specificity of the test data.


\section{Conclusion}
We develop a framework for inference of the ROC curve that quantifies the uncertainty of the algorithm's performance with regard to the test data sensitivity and specificity. In particular, we construct confidence intervals with fixed sensitivity or specificity for every point on the ROC curve. Besides the setting where we have test data from the same population as the observed data, we also consider the case where we have non-iid test data, and the uncertainty quantification is tailored to the shifted distribution. We also briefly discuss how to quantify ROC uncertainty for more general multi-label classification problems. The coverage is guaranteed both theoretically and numerically with numerical simulations.

\end{document}